# Collaborative planning and optimization for electric-thermal-hydrogen-coupled energy systems with portfolio selection of the complete hydrogen energy chain

Xinning Yi[a], Tianguang Lu[a,b]*, Yixiao Li[c], Qian Ai[d], Ran Hao[e]

[a]*School of Electrical Engineering, Shandong University, Jinan 250061, China*
[b]*School of Engineering and Applied Sciences and Harvard China Project, Harvard University, Cambridge, MA 02138, United States*
[c]*State Grid Shandong Electric Power Company Marketing Service Center (Metrology Center), Jinan 250013, China*
[d]*Key Laboratory of Control of Power Transmission and Conversion, Shanghai Jiao Tong University, Shanghai 200240, China*
[e]*National Power Dispatching and Control Center State Grid Coporation of China, Beijing 100000, China*

---

**Abstract**

Under the global low-carbon target, the uneven spatiotemporal distribution of renewable energy resources exacerbates the uncertainty and seasonal power imbalance. Additionally, the issue of an incomplete hydrogen energy chain is widely overlooked in planning models, which hinders the complete analysis of the role of hydrogen in energy systems. Therefore, this paper proposes a high-resolution collaborative planning model for electricity-thermal-hydrogen-coupled energy systems considering both the spatiotemporal distribution characteristics of renewable energy resources and the multi-scale bottom-to-top investment strategy for the complete hydrogen energy chain. Considering the high-resolution system operation flexibility, this paper proposes a hydrogen chain-based fast clustering optimization method that can handle high-dimensional data and multi-time scale operation characteristics. The model optimizes the geographical distribution and capacity configuration of the Northeast China energy system in 2050, with hourly operational characteristics. The planning optimization covered single-energy devices, multi-energy-coupled conversion devices, and electric-hydrogen transmission networks. Last but not least, this paper thoroughly examines the optimal portfolio selection of different hydrogen technologies based on the differences in cost, flexibility, and efficiency. In the Pareto analysis, the proposed model reduces CO2 emissions by 60% with a competitive cost. This paper provides a zero-carbon pathway for multi-energy systems with a cost 4% less than the social cost of carbon $44.6/ton, and the integration of the complete hydrogen energy chain reduces the renewable energy curtailment by 97.0%. Besides, the portfolio selection results indicate that the system favors the SOEC with the highest energy efficiency and the PEMFC with the fastest dynamic response when achieving zero-carbon emissions.

---

**Nomenclature**

***Abbreviations***

| | |
|---|---|
| AEC | alkaline electrolysis cell |
| BES | battery energy storage |

* Corresponding author.
*E-mail address:* tlu@sdu.edu.cn (T. Lu).

| | |
|---|---|
| CF | capacity factor |
| CHP | combined heat and power |
| COP | compressor |
| EB | electric boiler |
| EC | electrolysis cell |
| EIM | East Inner Mongolia |
| ES | energy storage |
| FC | fuel cell |
| GESS | gas energy storage system |
| HLJ | Heilongjiang |
| HPS | hydroelectric pumped storage |
| HS | hydrogen storage |
| HST | heat storage tank |
| HT | hydrogen-fired turbine |
| IES | integrated energy system |
| JL | Jilin |
| L | large |
| LN | Liaoning |
| M | medium |
| MCFC | molten carbonate fuel cell |
| P2G | power-to-gas |
| PAFC | phosphoric acid fuel cell |
| PEMEC | proton exchange membrane electrolysis cell |
| PEMFC | proton exchange membrane fuel cell |
| PV | photovoltaic panel |
| S | small |
| SCC | social cost of carbon |
| SCR | selective catalytic reduction |
| SOEC | solid oxide electrolysis cell |
| SOFC | solid oxide fuel cell |
| TU | thermal unit |
| WT | wind turbine |

***Parameters***

$\alpha$ unit consumption

$\beta$ energy conversion rate

$\eta$ efficiency

$\underline{\mu} / \overline{\mu}$ minimum/maximum output ratios

$\Phi$ slack variable for ES

$\Gamma$ RPS requirement

$a$ amortized investment cost

$c$ variable operating cost

$CF$ capacity factor

$ec$ emission factor

$E$ forecasting errors

$f$ fixed O&M cost

$l$ length of hydrogen pipeline

$LHV$ low heating value of hydrogen

$N$ number

$p$ unit price

$r^u / r^d$ ratios of upward/downward ramping

$re$ reserve capacity

$R^u / R^d$ start-up/shut-down ramp limits

$T$ time period

***Variables***

$\chi$ objective function of hydrogen transmission

$C$ cost

$E$ earning

$EC$ $CO_2$ emission

$H$ heat power

$\underline{I} / I / \overline{I}$ existing/newly/total installed capacity

$\underline{L} / L / \overline{L}$ existing/newly/total transmission capacity

$\underline{\mu} / \overline{\mu}$ transmission line capacity

$m$ hydrogen energy

| | |
|---|---|
| $M$ | hydrogen energy capacity |
| $O$ | online capacity |
| $P$ | electricity power |
| $Q$ | total hydrogen production |
| $R$ | revenue |
| $soc$ | state of charge |
| $S$ | shut-down capacity |
| $U$ | start-up capacity |
| ***Subscripts and superscripts*** | |
| ch | charging of storage |
| curt | curtailment |
| dis | discharging of storage |
| D | demand |
| e | electricity |
| e2h | electricity to hydrogen |
| es | energy-specific |
| h | heat |
| $H_2$ | hydrogen |
| lim | limit |
| loss | loss |
| o | oxidation |
| ps | power-specific |
| r | region |
| st | start-up |
| total | total |
| w | water |
| ***Indices and sets*** | |
| $i$ | technology |
| $j,k$ | region j to region k |
| $k$ | region |
| $t$ | time |
| X | set of equipment types (EC/HT/FC) |
| $\Psi$ | set of transmission line/pipeline |

## 1. Introduction

Under the global goal of carbon-neutral development, the vigorous deployment of clean and low-carbon renewable energy has become a vital way to deepen the decarbonization of the world's energy industry. China, as the world's largest $CO_2$ producer, proposed a series of policies to promote the development of renewable energy. According to statistics, in 2021, China's installed capacity of wind energy is 329.0 GW and solar energy is 306.4 GW, with average annual growth rates of 21.6% and 58.3% respectively in the past ten years [1]. It is expected that by 2050, China's installed wind energy capacity will exceed 2400 GW and solar energy capacity will exceed 2700 GW, with renewable energy power generation accounting for 85.8% of total power generation [2]. However, the randomness and volatility brought by high renewable penetration make energy systems face serious challenges in terms of flexibility and safe operation. Furthermore, the uneven spatiotemporal distribution of renewable energy resources in China intensifies the problem of imbalanced geographical development and insufficient flexibility of the energy system. Limited renewable energy consumption led to the phenomenon of seriously abandoning wind and solar power. Therefore, there is an increasingly urgent need for energy systems to spatiotemporally coordinate the allocation of renewable energy resources and achieve efficient renewable energy consumption.

As a high-density energy source with the advantages of flexible storage and conversion, high combustion calorific value, low carbon, and cleanliness, hydrogen is regarded as a crucial carrier to realize low-carbon energy transformation. Developing the hydrogen industry helps the energy system adapt to the future development trend of a higher renewable penetration rate [3]. Heating, as an important part of energy consumption, accounted for almost 50% of global end-use energy consumption in 2021, and CO2 emissions from heating consumption accounted for 39% of total $CO_2$ emissions [4]. In Northern China, winter heating is highly dependent on combined heat and power (CHP) units, and the operational characteristics of CHP units, which are determining-power-by-heat, severely limit the flexibility of the energy system and exacerbate the wind curtailment phenomenon [5]. The use of renewable energy for the electrification of heating has become a significant way to promote $CO_2$ emission reduction and enhance the energy system's consumption capacity [6]. In addition, waste heat utilization of hydrogen energy equipment such as electrolysis cells (EC), fuel cells (FC), and hydrogen-fired turbines (HT) can also be utilized for heat supply [7]-[9]. Therefore, exploring the planning and optimization of energy systems with electric-thermal-hydrogen-coupled is of great practical significance to coordinate the continuously growing demand for electricity and heat as well as to improve energy utilization.

In recent years, more attention has been given to the research of multi-energy-coupled energy system planning. Existing research mainly takes the optimal environmental or economic benefits as the optimization objective and solves the single/multi-objective multi-energy-coupled system collaborative planning model with the solution method of hierarchical optimization or global optimization. Yuchen Pu et al. [10] proposed a model of island integrated energy system (IES) and used a life cycle cost scheme considering system degradation to analyze the overall economy of the IES with power-hydrogen-heat-cooling cogeneration. Literature [11] proposed a planning approach for an electric-thermal-coupled IES and used a residential community as an example to verify the $CO_2$ reduction effect of hydrogen storage (HS). Based on electric-thermal-coupled, a planning configuration model of a net-zero energy building in the California region was designed [12], and the results showed that the utilization of natural gas fuel cells reduces fuel consumption and $CO_2$ emissions. Xiaojun Shen et al. [13] designed a multi-energy integrated microgrid scheme with HS to realize zero carbon emission in renewable-rich remote areas. Lukas Weimann et al. [14] designed a zero-carbon-emission electricity-hydrogen-coupled energy system led by wind energy and optimized the capacity and configuration of EC, FC, and HS equipment in the Netherlands. However, the economic benefits of hydrogen waste heat utilization and hydrogen production byproducts were not fully considered in the above studies, underestimating the efficiency of hydrogen utilization in the energy system. Meanwhile, most studies overlooked the impact of renewable spatiotemporal characteristics on the flexibility of the energy systems and did not consider the geographic distribution of the planning allocations.

The participation of hydrogen energy as a crucial energy carrier in the energy system is increasing, specifically in the links of production, compression, storage, transportation, and application [15]. Globally, many countries have adopted the hydrogen energy chain as an important component of their energy development strategies, with Japan and Germany as notable examples. Japan expects to establish an integrated upstream and downstream international hydrogen energy supply chain by 2030, realizing a 75% reduction in the cost of EC, the construction of pure hydrogen-fired power plants, and the deployment of 800,000 fuel cell vehicles[16]. Germany's updated National Hydrogen Strategy, released in July 2023, sets out 68 initiatives targeting the development of the entire hydrogen value chain, including doubling electrolysis capacity to 10 GW and expanding 1,800 kilometers of hydrogen pipelines by 2030 [17].

Current research mostly focuses on the analysis of specific links in the hydrogen energy chain. For the production link, Shaojie Song et al. [18] explored the role of green hydrogen produced from the electrolysis of water powered by renewables in Indian energy system planning, with emphasis on green hydrogen substitution for gray hydrogen to promote deep decarbonization in the industrial and transportation sectors. For the compression link, the literature [19] established a refined hydrogen COP model with different pressure ratios, and the optimization planning results showed that the model can reduce the installed COP capacity. For the storage link, Samira S. Farahani et al. [20] utilized HS in salt caverns as an alternative to large-scale battery energy storage, effectively reducing the cost of the IES by approximately 72.40% in 2050, with approximately 98.32% of the cost reduction coming from energy storage. Guangsheng Pan et al. [21] proposed a planning model for an electricity-hydrogen-coupled energy system considering hydrogen production from renewable energy sources and demonstrated the role of HS for seasonal energy complementation. For the transportation link, the literature [22] used hydrogen buffer tanks and hydrogen transport networks to construct a hydrogen energy supply chain, which was jointly planned and optimized in conjunction with the ammonia industry and the power grid. The application link is usually analyzed in conjunction with the production and storage links. Literature [23] proposed a gas energy storage system (GESS) combining power-to-gas (P2G) technology with HT, while literature [24] and [25] considered EC, HS, and FC capacity configuration in the planning stage. The above studies analyzed the role of hydrogen energy in the energy system, but only some of the links were involved and no complete hydrogen energy chain was formed. Furthermore, most of them ignored the waste heat utilization of hydrogen energy equipment, hydrogen production byproducts, and hydrogen pipeline planning. Considering the size, efficiency, and cost of EC technologies, Peng Hou et al. [26] performed equipment selection analysis for EC and optimized operations based on different planning schemes. In addition to cost and efficiency, literature [27] took into account flexibility differences to analyze the impact of EC technologies portfolio selection on the operating cost of the energy system and consumption rates of renewable energy. However, the above studies lack analyses on the portfolio selection of hydrogen energy equipment other than EC. In summary, there is a lack of multi-energy system planning models that take into account a comprehensive and quantitative analysis of both hydrogen energy links and equipment technology types. This requires the optimization model to consider the planning of EC, COP, HS, HT, FC, and hydrogen pipelines, combination sizing, and waste heat utilization while covering the production, compression, storage, transportation, and application links.

Therefore, the objective of this study is to develop a high-resolution planning model to optimize the geographic allocation of renewable investments and analyze the details about the role of the full hydrogen energy chain in the future energy system. The impacts of RPS, different technology types, and waste heat utilization on system planning costs and $CO_2$ reduction are analyzed. The main contributions of this work are summarized as follows:

1) Considering multiple equipment and energy conversion forms, we establish a refined comprehensive model of the complete hydrogen energy chain and integrate it with energy systems for collaborative planning of electricity, heat, and hydrogen based on the spatiotemporal characteristics of renewable energy resources;

2) Considering electric-hydrogen network expansion planning and establishing flexibility constraints using a hydrogen chain-based fast clustering technique for efficient long-term high-resolution simulation of coupled multi-energy flow interactions across space and time;

3) Conducting portfolio selection analysis of hydrogen energy equipment to determine selection laws based on performance differences on various technologies in terms of economy, flexibility, and efficiency.

The structure of the paper is organized as follows: Section 2 describes the system framework, the computational methods for the pre-data, and the model structure. A case discussion and an analysis of the relevant results are presented in Section 3, and conclusions are summarized in Section 4.

## 2. Methodology

### *2.1. Framework*

In this study, the research framework combines the spatiotemporal distribution characteristics of renewable energy resources with the planning of an electricity-thermal-hydrogen-coupled multi-regional energy system considering the complete hydrogen energy chain and equipment portfolio selection, which is illustrated in Fig. 1. The wind and solar hourly capacity factors are evaluated based on high-resolution hourly wind speed, solar irradiance and temperature data derived from a NASA meteorological database. The hourly heat demand is calculated based on temperature data, population density, and historical heating capacity data. In addition to the above data, the data used as input to the proposed optimization model include the regional topology, electricity demand, hydrogen demand, fossil fuel price, technical parameters of various energy conversion and storage equipment (e.g. current installed capacity, cost, efficiency), and transmission network parameters. The model takes the optimal total system cost or $CO_2$ emissions as the objective function and performs hourly balancing of different energy forms in each region based on fulfilling the energy demand of the target year. The output includes cost, $CO_2$ emissions, optimal technology planning and geographic distribution, hourly dispatch results, and transmission network structure.

The proposed multi-energy system model involves three forms of energy (electricity, heat, and hydrogen) and the interactions between them. In the electricity sector, an electricity system capacity expansion model from the literature [28] is used as a basis, including conventional thermal units (TU), wind turbines (WT), photovoltaic panels (PV), conventional energy storage (ES) technologies, electrical load demand, and electricity transmission lines. Based on this model, a transmission line expansion model is added and the hydrogen and thermal sectors are integrated. In addition to the base model generation technologies, the power balance also takes into account the electrical loads generated by the operation of the electric boiler (EB), EC, and compressor (COP), as well as the power output of the hydrogen generation units. In the thermal sector, heat demand is mainly met by CHP units and electric boilers (EB), and regulated by heat storage tanks (HST). According to the nature of heat energy, this paper only considers the internal balance of the region and does not consider regional heat transfer [29]. In the hydrogen sector, a complete hydrogen energy chain and equipment portfolio selection model is established. The complete hydrogen energy chain includes production, compression, storage, transportation, and application links, taking into account hydrogen energy technologies such as EC, COP, HS, HT, FC, and hydrogen pipelines. Equipment portfolio selection involves different technology types for EC, HT,

FC, and HS. Besides, waste heat utilization from hydrogen conversion equipment is used to meet the heat load demand.

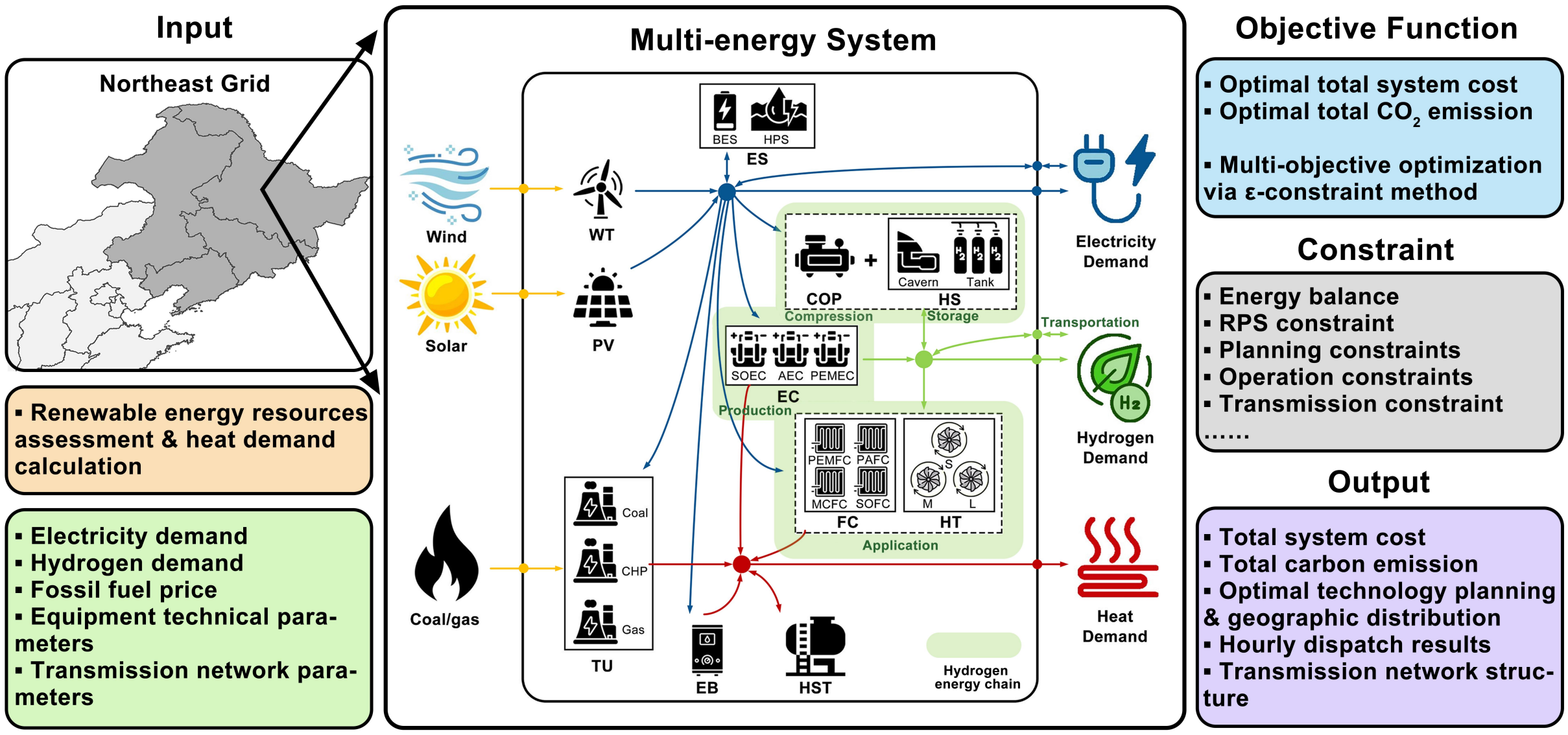

Fig. 1. Research framework

## *2.2. Renewable energy resources assessment and heat demand calculation*

The spatiotemporal distribution characteristics of regional renewable energy resources and unit operating characteristics affect renewable energy planning and geographic distribution. Therefore, this paper utilizes the methodologies of the literature [30] and [31] to evaluate the hourly capacity factors of regional wind and solar energy, respectively. The hourly wind speed, solar radiation, and temperature data used in this study are derived from the NASA meteorological database MERRA with a spatial resolution of 0.5 longitude by 0.667 latitude. The wind capacity factor calculation is performed using a GoldWind 1.5 MW turbine with a 100 m hub height, which requires the application of the power-law vertical wind speed profile formula [30] to convert the 50 m hourly wind speed data to 100 m hourly wind speed data. The output power of the WT and the grid unit wind energy per hour capacity factor is calculated based on the formula [30] for typical WT power. The grid unit solar energy per hour capacity factor is calculated based on the solar energy capacity factor generated by solar PV panels under the maximum power tracking method equation [31]. Considering the limitations of different land use types and slopes [28], geographic areas unsuitable for equipment deployment were filtered out for wind and solar energy using land use data from the MODIS database and altitude and slope data from the SRTM30 database. Combined with the regional topology and filtered data, hourly and average annual capacity factors for wind and solar were calculated for each region.

Heat load demand has a significant seasonal feature and is mainly determined by the difference between indoor and outdoor temperatures. In this study, the hourly heat load demand for each region is calculated using the methodology in reference [29], which combines the regional topology, temperature, population density, and regional hot water heating data. The calculation process uses 18 ℃ as the standard indoor temperature, and the outdoor ambient temperature is calculated based on the surface temperature data from the MERRA database and the population density data from the GPW database. The regional hot water heating data include the total annual regional hot water heating and the hot water heating capacity obtained from the literature [32].

*2.3. Hydrogen chain-based fast clustering optimization method*

A premise for the refined analysis of the hydrogen energy chain and equipment portfolio selection is the requirement of a unified operation model containing detailed flexibility constraints. The structure of large-scale EC and FC applied in energy systems is shown in Fig. 2 by combining low-power electric stacks into modules and further composing multiple modules into high-power equipment clusters. Each module unit in the cluster can be independently controlled to adjust the start/stop state and load power. In this paper, hydrogen chain equipment such as EC, FC, and HT are clustered by different technology types. For the large-scale medium- and long-timescale planning problem, if the 0/1 variables are used to represent the start/stop state of certain units/modules according to the traditional flexibility operational constraints, a large number of discrete decision variables will be introduced, leading to a computationally infeasible model.

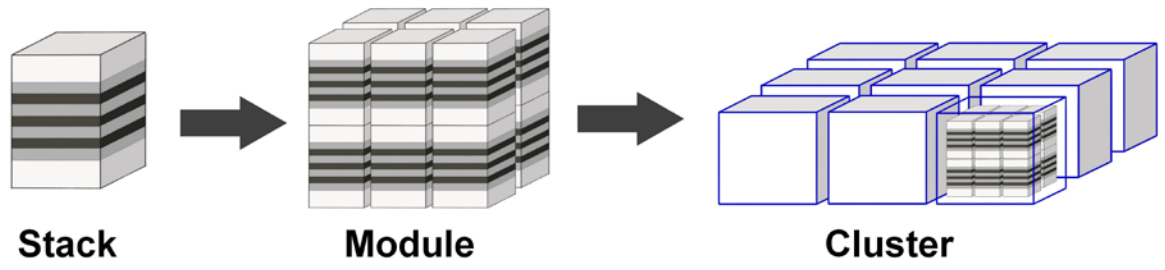

Fig. 2. EC and FC clustering composition

The fast cluster optimization method first optimizes the cluster after clustering multiple similar devices as a whole instead of considering individual modules and introduces three types of continuous variables, O, U, and S, to simulate the capacity temporal change of a certain cluster, as shown in Fig. 3.

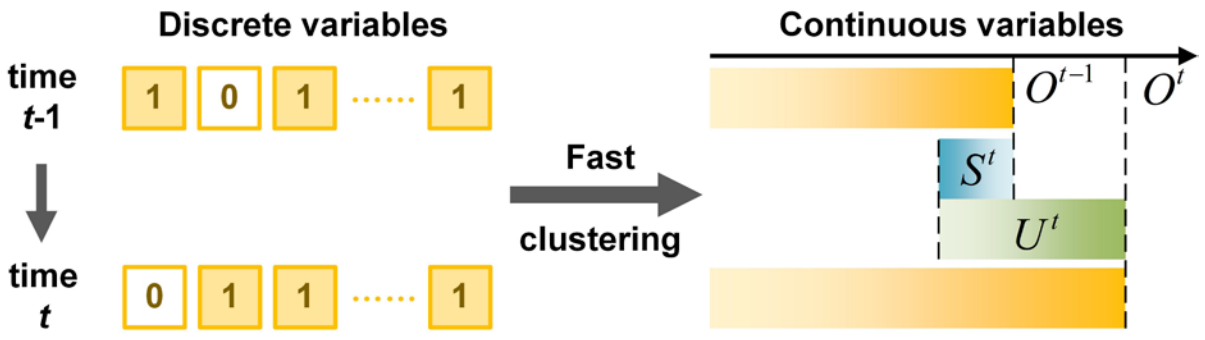

Fig. 3. Schematic diagram illustrating the temporal variation of cluster online capacity

$O^t$ , $U^t$ , and $S^t$ denote the online operating capacity, startup capacity, and shutdown capacity of a particular cluster at time t, respectively. All three should be less than the sum of the capacities of all modules of the cluster, and their logical relationships are as follows:

$$O^t - U^t = O^{t-1} - S^t \tag{1}$$

After the fast clustering optimization method, the maximum possible error of the cluster at a certain moment in time is the minimum capacity of the modules in the cluster, and the error impact decreases with the increase in the cluster's total capacity. The error impact is negligible for large-scale energy system planning in the medium and long term.

### 2.4. Complete hydrogen energy chain model

Under a high renewable penetration rate, the integration of hydrogen into energy systems can contribute to increased system flexibility and reduced renewable energy curtailment. This paper utilizes a hydrogen chain-based fast clustering optimization method to propose a refined model of the full hydrogen energy chain, which contains the capacity planning and hourly operation of related equipment and involves production, compression, storage, transportation, and application, as illustrated in Fig. 4.

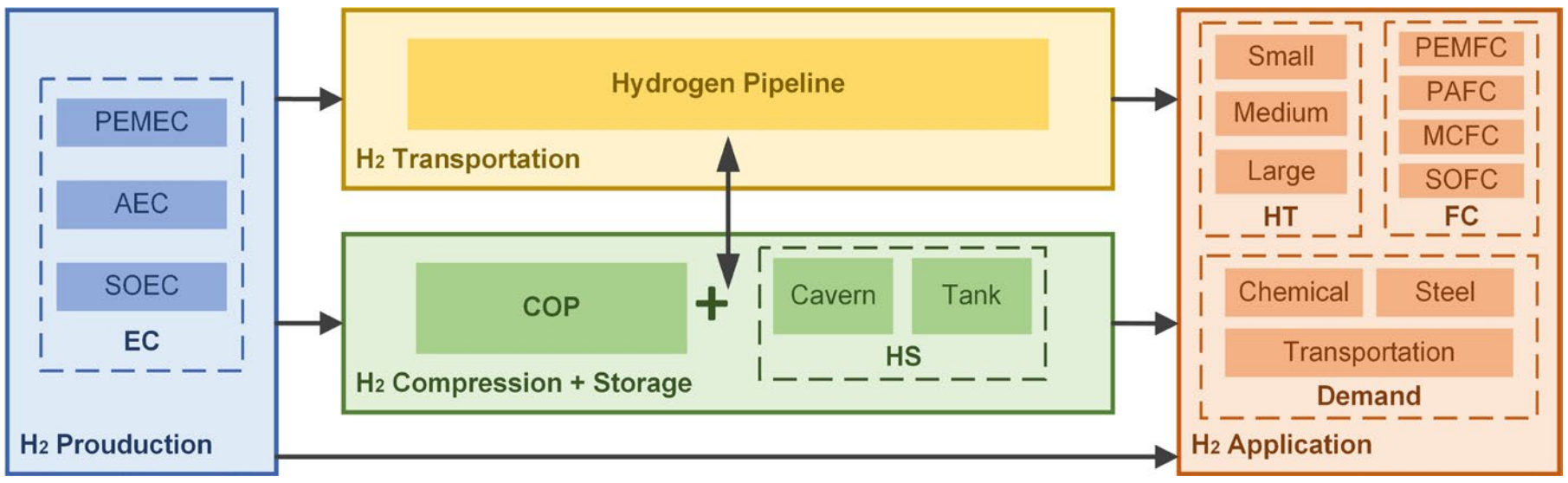

Fig. 4. Links composed of hydrogen energy chain

To promote the system's renewable energy consumption and fulfil low-carbon goals, the EC in the proposed hydrogen chain takes advantage of surplus wind and solar power for green hydrogen production. The hydrogen generated can serve as a feedstock to meet the demand for green hydrogen in the chemical, steel, and transportation industries in the region. Considering the time-varying and uncertain character of renewable energy sources, HS equipment combined with COP is utilized to store and release hydrogen at different times to regulate the balance between hydrogen supply and demand. In addition to satisfying hydrogen demand directly, hydrogen application includes hydrogen-electricity conversion using HT or FC for clean power supply. The uneven spatial distribution of renewable energy sources leads to a large gap in the potential of hydrogen that can be produced in different regions. Hence, this study utilizes hydrogen pipelines for inter-regional hydrogen transfer to optimize regional energy allocation. Furthermore, to promote hydrogen-thermal energy coupling, the hydrogen energy chain model introduces the utilization of waste heat from EC, HT, and FC to supplement the supply of heat demand.

#### 2.4.1. EC model

As the basis of the production link, the EC is the crucial equipment in the hydrogen energy chain, which is based on the principle of ionizing water with electricity to produce pure $H_2$ and oxygen as a byproduct. Three types of EC technologies are considered in this paper, AEC, PEMEC, and SOEC, with the technical parameters and cost data shown in Table 1, and their common basic model is:

$$0 \le \sum_{i=1}^{N_{\mathrm{EC}}} P_{\mathrm{EC},i}^{k,t} \le CF_{\mathrm{WT}}^{k,t} \cdot \overline{I}_{\mathrm{WT}}^{k} - P_{\mathrm{WT}}^{k,t} + CF_{\mathrm{PV}}^{k,t} \cdot \overline{I}_{\mathrm{PV}}^{k} - P_{\mathrm{PV}}^{k,t} \tag{2}$$

$$\beta_{e2h} = 3600 / LHV \tag{3}$$

$$m_{\mathrm{EC},i}^{k,t} = \beta_{e2h} \cdot \eta_{\mathrm{EC},i}^{\mathrm{e}} \cdot P_{\mathrm{EC},i}^{k,t} \tag{4}$$

$$H_{\mathrm{EC},i}^{k,t}=\eta_{\mathrm{EC},i}^{\mathrm{h}}\cdot(1-\eta_{\mathrm{EC},i}^{\mathrm{e}})\cdot P_{\mathrm{EC},i}^{k,t} \tag{5}$$

$$0\le Q=\sum_{k=1}^{N_{\mathrm{r}}}\sum_{i=1}^{N_{\mathrm{EC}}}\sum_{t=1}^{T}\left(\beta_{e2h}\cdot\eta_{\mathrm{EC},i}^{\mathrm{e}}\cdot P_{\mathrm{EC},i}^{k,t}\right) \tag{6}$$

$$C_{\mathrm{w}}=\alpha_{\mathrm{EC}}^{\mathrm{w}}\cdot p_{\mathrm{w}}\cdot Q \tag{7}$$

$$E_{\mathrm{o}}=\alpha_{\mathrm{EC}}^{\mathrm{o}}\cdot p_{\mathrm{o}}\cdot Q \tag{8}$$

$$E_{\mathrm{H_2}}=p_{\mathrm{H_2}}\cdot\sum_{k=1}^{N_{\mathrm{r}}}\sum_{t=1}^{T}m_{D}^{k,t} \tag{9}$$

*2.4.2. HT model*

HT is an essential energy device to promote the coupling of electricity-thermal-hydrogen, its technical parameters and basic model are similar to those of gas-fired units, and the cost difference between the two is mainly from the addition of selective catalytic reduction (SCR) equipment for reducing the combustion emission of nitrogen oxides (NOx) [8]. The above specific data are shown in Table 1. In addition to the table data, HT has a variable operating cost of \$0.002/kWh [8]. In this paper, the HT is mainly categorized into small (S), medium (M), and large (L) size units, and the basic model is:

$$P_{\mathrm{HT},i}^{k,t}=\eta_{\mathrm{HT},i}^{\mathrm{e}}\cdot m_{\mathrm{HT},i}^{k,t}/\beta_{e2h} \tag{10}$$

$$H_{\mathrm{HT},i}^{k,t}=\eta_{\mathrm{HT},i}^{h}\cdot\left(1-\eta_{\mathrm{HT},i}^{e}\right)\cdot m_{\mathrm{HT},i}^{k,t}/\beta_{e2h} \tag{11}$$

*2.4.3. FC model*

FC is hydrogen power generation equipment, which promotes multi-energy synergy in the energy system, the same as HT. The basic principle of FC is to generate electrical energy through the movement of electrons during redox reactions and to recover thermal energy through heat collectors to realize cogeneration. FC is a clean power generation technology with potential applications in the future, and there is a rich variety of them. In this paper, we select four technologies suitable for distributed generation in the energy system, namely PEMFC, PAFC, MCFC, and SOFC, with the technical parameters and cost data listed in Table 1, and their common basic model is:

$$P_{\mathrm{FC},i}^{k,t}=\eta_{\mathrm{FC},i}^{\mathrm{e}}\cdot m_{\mathrm{FC},i}^{k,t}/\beta_{e2h} \tag{12}$$

$$H_{\mathrm{FC},i}^{k,t}=\eta_{\mathrm{FC},i}^{h}\cdot\left(1-\eta_{\mathrm{FC},i}^{e}\right)\cdot m_{\mathrm{FC},i}^{k,t}/\beta_{e2h} \tag{13}$$

*2.4.4. HS and COP model*

HS as seasonal energy storage can balance the fluctuation of supply and demand across time scales and is the key equipment to realize the medium- and long-term utilization of hydrogen energy and the coordination of multi-energy. The hydrogen density is low, and the utilization of COP can significantly reduce the storage volume. In this paper, the COP is connected to the HS, and the charging/discharging of hydrogen is limited by the power of the COP. We considered two hydrogen storage technologies: hydrogen storage tanks and salt cavern hydrogen storage. Their technical parameters are similar, the cost difference is large, and the specific data are summarized in Table 1. Moreover, the loss rates for both HS and COP are taken to be zero [38]. The basic model of HS and COP connection is:

$$0 \le m_{\mathrm{HS},i}^{\mathrm{dis},k,t}, m_{\mathrm{HS},i}^{\mathrm{ch},k,t} \le M_{\mathrm{COP},i}^{k} \tag{14}$$

$$0 \le soc_{\mathrm{HS},i}^{k,t} \le M_{\mathrm{HS},i}^{k,t} \tag{15}$$

$$0 \le soc_{\mathrm{HS},i}^{k,t} - m_{\mathrm{HS},i}^{\mathrm{dis},k,t} \tag{16}$$

$$soc_{\mathrm{HS},i}^{k,t+1} - soc_{\mathrm{HS},i}^{k,t} = \eta_{\mathrm{HS},i}^{\mathrm{ch}} \cdot m_{\mathrm{HS},i}^{\mathrm{ch},k,t} - m_{\mathrm{HS},i}^{\mathrm{dis},k,t} / \eta_{\mathrm{HS},i}^{\mathrm{dis}} - \eta_{\mathrm{HS},i}^{\mathrm{loss}} \cdot soc_{\mathrm{HS},i}^{k,t} \tag{17}$$

$$soc_{\mathrm{HS},i}^{k,1} = soc_{\mathrm{HS},i}^{k,T} \tag{18}$$

$$P_{\mathrm{COP},i}^{k,t} = \alpha_{\mathrm{COP}}^{\mathrm{e}} \cdot \left( m_{\mathrm{HS},i}^{\mathrm{ch},k,t} + m_{\mathrm{HS},i}^{\mathrm{dis},k,t} \right) \tag{19}$$

*2.4.5. Hydrogen pipeline model*

Since only the hydrogen demand data for the whole region are available, this paper analyzes the role of hydrogen transmission between regions by performing the planning of hydrogen pipelines in the post-processing optimization. The specific process is as follows: considering the overall hydrogen balance in the Northeast region in the model planning simulation, ignoring the inter-regional hydrogen pipeline transportation limitations, and solving the pipeline planning separately after obtaining the results of equipment planning and configuration. It is assumed that new hydrogen pipelines are possible between adjacent regions. The minimum value of the sum of the products of the hydrogen transmission volume and the length of each pipeline in the region is taken as the objective function and is modelled as follows:

$$\chi = \sum_{j \in \Psi_{\mathrm{H}_2}^{k}} \sum_{t=1}^{T} l_{j,k} \cdot \left| m_{j,k}^{t} \right| \tag{20}$$

$$\sum_{j \in \Psi_{\mathrm{H}_2}^{k}} m_{j,k}^{t} + \sum_{i=1}^{N_{\mathrm{EC}}} m_{\mathrm{EC},i}^{k,t} + \sum_{i=1}^{N_{\mathrm{HS}}} m_{\mathrm{HS},i}^{\mathrm{dis},k,t} - \sum_{i=1}^{N_{\mathrm{HT}}} m_{\mathrm{HT},i}^{k,t} - \sum_{i=1}^{N_{\mathrm{FC}}} m_{\mathrm{FC},i}^{k,t} - \sum_{i=1}^{N_{\mathrm{HS}}} m_{\mathrm{HS},i}^{\mathrm{ch},k,t} - m_{\mathrm{D}}^{k,t} = 0 \tag{21}$$

*2.4.6. Operational constraints of EC, HT and FC*

In this paper, a hydrogen chain-based fast clustering optimization method is applied to improve the operation constraints of EC, HT, and FC by clustering different technology types of equipment by region and introducing linear continuous variables to describe the start-stop constraints and ramping constraints, the details of which are presented as follows:

$$0 \le P_{\mathrm{X},i}^{k,t} \le O_{\mathrm{X},i}^{k,t} \le \overline{I}_{\mathrm{X},i}^{k} = I_{\mathrm{X},i}^{k} + \underline{I}_{\mathrm{X},i}^{k} \tag{22}$$

$$P_{\mathrm{X},i}^{k,t} - P_{\mathrm{X},i}^{k,t-1} = S_{\mathrm{X},i}^{k,t} - U_{\mathrm{X},i}^{k,t} \tag{23}$$

$$0 \le S_{\mathrm{X},i}^{k,t}, U_{\mathrm{X},i}^{k,t} \le \overline{I}_{\mathrm{X},i}^{k} \tag{24}$$

$$\underline{\mu}_{\mathrm{X},i} \cdot O_{\mathrm{X},i}^{k,t} \le P_{\mathrm{X},i}^{k,t} \le \overline{\mu}_{\mathrm{X},i} \cdot O_{\mathrm{X},i}^{k,t} \tag{25}$$

$$P_{\mathrm{X},i}^{k,t} \le \overline{\mu}_{\mathrm{X},i} \cdot \left( O_{\mathrm{X},i}^{k,t} - S_{\mathrm{X},i}^{k,t} - U_{\mathrm{X},i}^{k,t+1} \right) + R_{\mathrm{X},i}^{\mathrm{u}} \cdot S_{\mathrm{X},i}^{k,t} + R_{\mathrm{X},i}^{\mathrm{d}} \cdot U_{\mathrm{X},i}^{k,t+1} \tag{26}$$

$$\begin{cases} P_{\mathrm{X},i}^{k,t} - P_{\mathrm{X},i}^{k,t-1} \le r_{\mathrm{X},i}^{\mathrm{u}} \cdot \left( O_{\mathrm{X},i}^{k,t} - S_{\mathrm{X},i}^{k,t} \right) + R_{\mathrm{X},i}^{\mathrm{u}} \cdot S_{\mathrm{X},i}^{k,t} - \underline{\mu}_{\mathrm{X},i} \cdot U_{\mathrm{X},i}^{k,t} \\ P_{\mathrm{X},i}^{k,t-1} - P_{\mathrm{X},i}^{k,t} \le r_{\mathrm{X},i}^{\mathrm{d}} \cdot \left( O_{\mathrm{X},i}^{k,t} - S_{\mathrm{X},i}^{k,t} \right) - \underline{\mu}_{\mathrm{X},i} \cdot S_{\mathrm{X},i}^{k,t} + R_{\mathrm{X},i}^{\mathrm{d}} \cdot U_{\mathrm{X},i}^{k,t} \end{cases} \tag{27}$$

$$\begin{cases} 0 \le U_{\mathrm{X},i}^{k,t+1} \le O_{\mathrm{X},i}^{k,t} - \sum_{\tau=0}^{t-1} S_{\mathrm{X},i}^{k,t-\tau} & t = 1 \cdots T_{\mathrm{X},i}^{\mathrm{u}} - 1 \\ 0 \le U_{\mathrm{X},i}^{k,t+1} \le O_{\mathrm{X},i}^{k,t} - \sum_{\tau=0}^{T_{\mathrm{X},i}^{\mathrm{u}}-2} S_{\mathrm{X},i}^{k,t-\tau} & t = T_{\mathrm{X},i}^{\mathrm{u}} \cdots T - 1 \end{cases} \tag{28}$$

$$\begin{cases} 0 \le S_{\mathrm{X},i}^{k,t+1} \le \overline{I}_{\mathrm{X},i}^{k} - O_{\mathrm{X},i}^{k,t} - \sum_{\tau=0}^{t-1} U_{\mathrm{X},i}^{k,t-\tau} & t = 1 \cdots T_{\mathrm{X},i}^{\mathrm{d}} - 1 \\ 0 \le S_{\mathrm{X},i}^{k,t+1} \le \overline{I}_{\mathrm{X},i}^{k} - O_{\mathrm{X},i}^{k,t} - \sum_{\tau=0}^{T_{\mathrm{X},i}^{\mathrm{d}}-2} U_{\mathrm{X},i}^{k,t-\tau} & t = T_{\mathrm{X},i}^{\mathrm{d}} \cdots T - 1 \end{cases} \tag{29}$$

### *2.5. Collaborative planning optimization model*

#### *2.5.1. Objective function*

The objective function of the proposed model is to minimize the total system cost or the total $CO_2$ emissions. The total system cost consists of investment cost, O&M cost, and revenue, which are derived from the sale of hydrogen and the byproducts generated from hydrogen production.

$$C_{\mathrm{total}} = C_{\mathrm{TU}} + C_{\mathrm{WT}} + C_{\mathrm{PV}} + C_{\mathrm{ES}} + C_{\mathrm{EC}} + C_{\mathrm{HT}} + C_{\mathrm{FC}} + C_{\mathrm{HS}} + C_{\mathrm{COP}} + C_{\mathrm{EB}} + C_{\mathrm{HES}} + C_{\mathrm{L}} - R \tag{30}$$

$$C_{\mathrm{TU}}=\sum_{k=1}^{N_{\mathrm{r}}}\sum_{i=1}^{N_{\mathrm{TU}}}\left(a_{\mathrm{TU},i}\cdot I_{\mathrm{TU},i}+f_{\mathrm{TU},i}\cdot\overline{I}_{\mathrm{TU},i}+\sum_{t=1}^{T}\left(c_{\mathrm{TU},i}^{k}\cdot P_{\mathrm{TU},i}^{k,t}+c_{\mathrm{TU,i}}^{\mathrm{st}}\cdot S_{\mathrm{TU},i}^{k,t}\right)\right) \tag{31}$$

$$C_{\mathrm{WT}}=\sum_{k=1}^{N_{\mathrm{r}}}\left(a_{\mathrm{WT}}\cdot I_{\mathrm{WT}}^{k}+f_{\mathrm{WT}}\cdot\overline{I}_{\mathrm{WT}}^{k}\right) \tag{32}$$

$$C_{\mathrm{PV}}=\sum_{k=1}^{N_{\mathrm{r}}}\left(a_{\mathrm{PV}}\cdot I_{\mathrm{PV}}^{k}+f_{\mathrm{PV}}\cdot\overline{I}_{\mathrm{PV}}^{k}\right) \tag{33}$$

$$C_{\mathrm{ES}}=\sum_{k=1}^{N_{\mathrm{r}}}\sum_{i=1}^{N_{\mathrm{ES}}}\left(a_{\mathrm{ES},i}^{\mathrm{ps}}\cdot\overline{I}_{\mathrm{ES},i}^{\mathrm{ps},k}+a_{\mathrm{ES},i}^{\mathrm{es}}\cdot\overline{I}_{\mathrm{ES},i}^{\mathrm{es},k}+c_{\mathrm{ES},i}\cdot\sum_{t=1}^{T}\left(P_{\mathrm{ES},i}^{\mathrm{dis},k}+P_{\mathrm{ES},i}^{\mathrm{ch},k}\right)\right) \tag{34}$$

$$C_{\mathrm{EC}}=\sum_{k=1}^{N_{\mathrm{r}}}\sum_{i=1}^{N_{\mathrm{EC}}}\left(a_{\mathrm{EC},i}\cdot I_{\mathrm{EC},i}^{k}+f_{\mathrm{EC},i}\cdot\overline{I}_{\mathrm{EC},i}^{k}\right)+C_{\mathrm{w}} \tag{35}$$

$$C_{\mathrm{HT}}=\sum_{k=1}^{N_{\mathrm{r}}}\sum_{i=1}^{N_{\mathrm{HT}}}\left(a_{\mathrm{HT},i}\cdot I_{\mathrm{HT},i}^{k}+f_{\mathrm{HT},i}\cdot\overline{I}_{\mathrm{HT},i}^{k}+\sum_{t=1}^{T}\left(c_{\mathrm{HT},i}\cdot P_{\mathrm{HT},i}^{k,t}+c_{\mathrm{HT},i}^{\mathrm{st}}\cdot S_{\mathrm{HT},i}^{k,t}\right)\right) \tag{36}$$

$$C_{\mathrm{FC}}=\sum_{k=1}^{N_{\mathrm{r}}}\sum_{i=1}^{N_{\mathrm{FC}}}\left(a_{\mathrm{FC},i}\cdot I_{\mathrm{FC},i}^{k}+f_{\mathrm{FC},i}\cdot\overline{I}_{\mathrm{FC},i}^{k}\right) \tag{37}$$

$$C_{\mathrm{HS}}=\sum_{k=1}^{N_{\mathrm{r}}}\sum_{i=1}^{N_{\mathrm{HS}}}\left(a_{\mathrm{HS},i}\cdot M_{\mathrm{HS},i}^{k}+f_{\mathrm{HS},i}\cdot M_{\mathrm{HS},i}^{k}\right) \tag{38}$$

$$C_{\mathrm{COP}}=\sum_{k=1}^{N_{\mathrm{r}}}\sum_{i=1}^{N_{\mathrm{HS}}}\left(a_{\mathrm{COP}}\cdot M_{\mathrm{COP},i}^{k}+f_{\mathrm{COP}}\cdot M_{\mathrm{COP},i}^{k}\right) \tag{39}$$

$$C_{\mathrm{EB}}=\sum_{k=1}^{N_{\mathrm{r}}}\left(a_{\mathrm{EB}}\cdot I_{\mathrm{EB}}^{k}+f_{\mathrm{EB}}\cdot\overline{I}_{\mathrm{EB}}^{k}\right) \tag{40}$$

$$C_{\mathrm{HES}}=\sum_{k=1}^{N_{\mathrm{r}}}\left(a_{\mathrm{HES}}\cdot I_{\mathrm{HES}}^{k}+f_{\mathrm{HES}}\cdot\overline{I}_{\mathrm{HES}}^{k}+\sum_{t=1}^{T}c_{\mathrm{HES}}\cdot\left(P_{\mathrm{HES}}^{\mathrm{dis},k,t}+p_{\mathrm{HES}}^{\mathrm{ch},k,t}\right)\right) \tag{41}$$

$$C_{\mathrm{L}}=\sum_{j\in\Psi^{k}}a_{\mathrm{L}}^{k}\cdot L_{j,k} \tag{42}$$

$$R = E_{\text{o}} + E_{\text{H}_2} \tag{43}$$

Total $CO_2$ emissions are calculated from the $CO_2$ emission factor of a conventional TU and the TU's power generation.

$$EC_{\text{total}} = \sum_{k=1}^{N_{\text{r}}} \sum_{i=1}^{N_{\text{TU}}} \sum_{t=1}^{T} ec_{\text{TU},i}^{k} \cdot P_{\text{TU},i}^{k,t} \tag{44}$$

In this paper, we consider single-objective optimization or multi-objective optimization in conjunction with different scenarios and use the improved epsilon constraint method [39] to achieve Pareto-optimal multi-objective optimization.

*2.5.2. Constraints*

The constraints of the electricity sector in this paper are referenced in the literature [28], where three types of conventional TU are considered: coal-fired, CHP, and gas-fired units, and two types of conventional ES are considered: battery energy storage (BES) and hydroelectric pumped storage (HPS). Data on the technical parameters of conventional TU, conventional ES, WT, and PV are derived from the literature [34]. In addition, this paper considers the expansion of electricity transmission lines between different regions, which is modelled as follows:

$$-\overline{L}_{j,k} \leq P_{j,k} \leq \overline{L}_{j,k} \tag{45}$$

$$\overline{L}_{j,k} = \underline{L}_{j,k} + L_{j,k} \leq L_{j,k}^{\text{lim}} \tag{46}$$

The constraints for the hydrogen sector are described in detail in Section 2.3. For the thermal sector, the CHP unit adds a heat production constraint based on the literature [40], with a proportional relationship between the thermal and electrical efficiencies. The electrical efficiency of CHP is taken as 45%, and the thermal efficiency is taken as 30% [41].

$$H_{\text{CHP},i}^{k,t} = P_{\text{CHP},i}^{k,t} \cdot \eta_{\text{CHP},i}^{\text{h}} / \eta_{\text{CHP},i}^{\text{e}} \tag{47}$$

The EB generates heat with electricity, which can effectively reduce $CO_2$ emissions during the heating period, while the HST provides real-time heat storage/release for balancing heat supply and demand, which enhances the flexibility of the energy system. The technical parameters of both are summarized in Table 2. The relevant constraints are as follows:

$$0 \leq H_{\text{HST},k}^{\text{ch},k,t}, H_{\text{HST},k}^{\text{dis},k,t} \leq \overline{I}_{\text{HST}}^{\text{ps},k} \tag{48}$$

$$0 \leq soc_{\text{HST}}^{k,1} \leq \overline{I}_{\text{HST}}^{\text{es},k} \tag{49}$$

$$soc_{\text{HST}}^{k,1} = soc_{\text{HST}}^{k,T} \tag{50}$$

$$soc_{\mathrm{HST}}^{k,t+1}=soc_{\mathrm{HST}}^{k,t}+\eta_{\mathrm{HST}}^{\mathrm{ch}}\cdot H_{\mathrm{HST},k}^{\mathrm{ch},k,t}-H_{\mathrm{HST},k}^{\mathrm{dis},k,t}/\eta_{\mathrm{HST}}^{\mathrm{dis}}-\eta_{\mathrm{HST}}^{\mathrm{loss}}\cdot soc_{\mathrm{HST}}^{k,t} \tag{51}$$

This paper utilizes a renewable portfolio standard (RPS) to constrain the renewable energy penetration of energy systems:

$$\begin{aligned}&\sum_{k=1}^{N_{\mathrm{r}}}\sum_{t=1}^{T}\left(P_{\mathrm{WT}}^{k,t}+P_{\mathrm{WT}}^{k,t}+\sum_{i=1}^{N_{\mathrm{HT}}}P_{\mathrm{HT},i}^{k,t}+\sum_{i=1}^{N_{\mathrm{FC}}}P_{\mathrm{FC,i}}^{k,t}\right)\\&\geq\Gamma\cdot\sum_{k=1}^{N_{\mathrm{r}}}\left(\sum_{t=1}^{T}\left(P_{\mathrm{D}}^{k,t}+P_{\mathrm{ex}}^{k,t}+\sum_{i=1}^{N_{\mathrm{EC}}}P_{\mathrm{EC},i}^{k,t}+\sum_{i=1}^{N_{\mathrm{HS}}}P_{\mathrm{COP},i}^{k,t}+P_{\mathrm{EB}}^{k,t}\right)+\Phi_k\right)\end{aligned} \tag{52}$$

$$\Phi_k=\sum_{t=1}^{T}\sum_{i=1}^{N_{\mathrm{ES}}}\left(P_{\mathrm{ES},i}^{\mathrm{ch},k,t}-P_{\mathrm{ES},i}^{\mathrm{dis},k,t}\right) \tag{53}$$

The system reliability constraint is:

$$\begin{aligned}&\sum_{i=1}^{N_{\mathrm{TU}}}\bar{\mu}_{\mathrm{UT},i}\cdot O_{UT,i}^{k,t}+\alpha_k^t\cdot\bar{I}_{\mathrm{WT}}^k+\beta_k^t\cdot\bar{I}_{\mathrm{PV}}^k+\sum_{i=1}^{N_{\mathrm{ES}}}\left(P_{\mathrm{ES},i}^{\mathrm{dis},k,t}+re_{\mathrm{ES},i}^t\right)+\sum_{j\in\Psi^k}P_{j,k}^t-P_{\mathrm{ex}}^{k,t}+\sum_{i=1}^{N_{\mathrm{HT}}}P_{\mathrm{HT},i}^{k,t}\\&+\sum_{i=1}^{N_{\mathrm{FC}}}P_{\mathrm{FC},i}^{k,t}\geq P_{\mathrm{D}}^{k,t}+re_{\mathrm{d}}^t+e_{\mathrm{WT}}\cdot P_{\mathrm{WT}}^{k,t}+e_{\mathrm{PV}}\cdot p_{\mathrm{PV}}^{k,t}+\sum_{i=1}^{N_{\mathrm{EC}}}P_{\mathrm{EC},i}^{k,t}+\sum_{i=1}^{N_{\mathrm{HS}}}P_{\mathrm{COP},i}^{k,t}+P_{\mathrm{EB}}^{k,t}\end{aligned} \tag{54}$$

The balance constraints for the different energy forms of the energy systems are:

$$\begin{aligned}&\sum_{i=1}^{N_{\mathrm{TU}}}P_{UT,i}^{k,t}+P_{\mathrm{WT}}^{k,t}+p_{\mathrm{PV}}^{k,t}+\sum_{j\in\Psi^k}P_{j,k}^t-P_{\mathrm{ex}}^{k,t}+\sum_{i=1}^{N_{\mathrm{ES}}}\left(P_{\mathrm{ES},i}^{\mathrm{dis},k,t}-P_{\mathrm{ES},i}^{\mathrm{ch},k,t}\right)+\sum_{i=1}^{N_{\mathrm{HT}}}P_{\mathrm{HT},i}^{k,t}\\&+\sum_{i=1}^{N_{\mathrm{FC}}}P_{\mathrm{FC},i}^{k,t}=P_{\mathrm{D}}^{k,t}+\sum_{i=1}^{N_{\mathrm{EC}}}P_{\mathrm{EC},i}^{k,t}+\sum_{i=1}^{N_{\mathrm{HS}}}P_{\mathrm{COP},i}^{k,t}+P_{\mathrm{EB}}^{k,t}\end{aligned} \tag{55}$$

$$\sum_{i=1}^{N_{\mathrm{CHP}}}H_{\mathrm{CHP},i}^{k,t}+\sum_{i=1}^{N_{\mathrm{EC}}}H_{\mathrm{EC},i}^{k,t}+\sum_{i=1}^{N_{\mathrm{HT}}}H_{\mathrm{HT},i}^{k,t}+\sum_{i=1}^{N_{\mathrm{FC}}}H_{\mathrm{FC},i}^{k,t}+H_{\mathrm{EB}}^{k,t}+H_{\mathrm{HST},i}^{\mathrm{dis},k,t}=H_{\mathrm{HST},i}^{\mathrm{ch},k,t}+H_{\mathrm{D}}^{k,t}+H_{\mathrm{curt}}^{k,t} \tag{56}$$

$$0\leq H_{\mathrm{curt}}^{k,t} \tag{57}$$

$$\sum_{i=1}^{N_{\mathrm{EC}}}m_{\mathrm{EC},i}^t+\sum_{i=1}^{N_{\mathrm{HS}}}m_{\mathrm{HS},i}^{\mathrm{dis},t}=\sum_{i=1}^{N_{\mathrm{HT}}}m_{\mathrm{HT},i}^t+\sum_{i=1}^{N_{\mathrm{FC}}}m_{\mathrm{FC},i}^t+\sum_{i=1}^{N_{\mathrm{HS}}}m_{\mathrm{HS},i}^{\mathrm{ch},t}+m_{\mathrm{D}}^t \tag{58}$$

## 3. Case study

### 3.1. Case data

Table 1. Key parameters of hydrogen energy chain-related equipment

| Parameter | EC | | | HT | | | FC | | | | HS | | COP |
|---|---|---|---|---|---|---|---|---|---|---|---|---|---|
| | SOEC | AEC | PEMEC | S | M | L | PEMFC | PAFC | MCFC | SOFC | Cavern | Tank | |
| Capital cost[a] | 750 | 450 | 550 | 708 | 608 | 524 | 660 | 928 | 1031 | 722 | 4.5 | 450 | 360 |
| Fixed O&M cost (%) | 3 | 5 | 5 | 2.6 | 2.5 | 2.4 | 5 | | | | 2 | | 2 |
| Start-up cost ($/kW) | / | | | 0.088 | | | / | | | | / | | / |
| Electricity efficiency (%) | 83.5 | 75 | 70.5 | 60 | | | 57 | 40 | 60 | 70 | 95 | | / |
| Waste heat utilization efficiency (%) | 80 | | | 80 | | | 90 | 80 | 80 | 80 | / | | / |
| Load range (%) | 0-100 | 20-100 | 5-100 | 25-100 | 40-100 | 40-100 | 5-100 | 20-100 | 0-100 | 0-100 | 0-100 | | 0-100 |
| Ramping limit (%/h) | 30 | 50 | 100 | 100 | 50 | | 100 | 50 | 30 | 30 | 100 | | 100 |
| Start-up/shut-down times (hours) | 2/2 | 1/1 | 1/1 | 1/1 | 4/2 | 6/12 | 1/1 | 2/2 | 5/5 | 8/8 | / | | / |
| Lifetime (years) | 20 | 25 | 15 | 25 | | | 20 | 15 | 10 | 25 | 30 | | 15 |
| Ref. | [33]-[35] | | | [8],[28] | | | [36]-[37] | | | | [14],[21],[38] | | [38] |

[a] The units of capital costs for different equipment vary as follows: the unit of EC,HT and FC is $/kW, the unit of HS is $/kg and the unit of COP is $/kg • h

The case study focuses on the collaborative planning of electric-thermal-hydrogen-coupled energy systems based on the Northeast China power grid, with 2050 as the planning target year. A one-year hourly operational simulation is incorporated into the planning model. The Northeast grid is a regional grid including Liaoning (LN), Jilin (JL), Heilongjiang (HLJ), and East Inner Mongolia (EIM) grids. The regional grid is abundant in wind energy resources, and the energy demand is characterized by significant seasonal differences owing to the winter heating period. The electricity demand and existing installed capacity data in the region in 2020 are summarized in Table 3, which are obtained from the National Energy Administration. Except for local electricity demand, this model considers export electricity demand. The annual hourly heat load demand curve for the Northeast energy system is calculated by the methodology in Section 2.2. Afterward, the annual electricity demand and heat demand for each region in the target year are obtained by calculating the electricity growth rate [45] and heat growth rate [46] for the period 2020-2050, respectively. The annual hydrogen demand data for the Northeast energy system in 2035 [47] is 680000 t, with steel, chemical industry, and transportation sections, and the regional hydrogen demand in 2050 is obtained by the hydrogen growth rate [48], divided equally among periods. The technical parameters of the existing transmission lines in 2020 and the target year expansion capacity [49] are summarized in Table 4. In this paper, it is assumed that inter-regional transmission congestion requirements are satisfied and that the network structure remains unchanged. The technical parameters and cost data for each piece of equipment are projected for 2050 and are summarized in Table 1-2. The investment costs for the newly installed equipment are amortized over the lifetime at a 7% interest rate. According to the National Development and Reform Commission (NDRC), there are large inter-regional differences in coal and natural gas prices, with coal ranging from $37.3/ton in EIM to $80.3/ton in LN and natural gas ranging from $0.1794/m3 in EIM to $0.2706/m3 in LN. It should be noted that this paper assumes a USD-CNY exchange rate of 6.8 CNY/USD. Furthermore, the prices of water, oxygen, and hydrogen are set to \$0.01/kg, \$0.04/kg, and \$2/kg, respectively, in 2050 [35].

Table 2. Key parameters of EB and HST

| Parameter | EB | HST |
|---|---|---|
| Capital cost (\$/kW) | 125 | 25 |
| Fixed O&M cost (%) | 1.2 | 2 |
| Electricity efficiency (%) | 99 | 90 |
| Load range (%) | 0-100 | 0-100 |
| Ramping limit (%/h) | 100 | 100 |
| Lifetime (years) | 20 | 20 |
| Ref. | [42]-[43] | [43]-[44] |

Table 3. Northeast grid generating unit mix and electricity demand in 2020 (GW)

| | | | EIM | HLJ | JL | LN |
|---|---|---|---|---|---|---|
| WT | | | 12.4 | 8.6 | 8.3 | 11.0 |
| PV | | | 1.0 | 4.2 | 3.7 | 5.0 |
| TU[a] | Coal | S | 0.6 | 0.7 | 0.5 | 1.1 |
| | | M | 3 | 3.5 | 2.7 | 5.6 |
| | | L | 7.9 | 9.1 | 7 | 14.6 |
| | CHP | S | 0.8 | 0.9 | 0.7 | 2.2 |
| | | M | 6.8 | 7.9 | 6.1 | 10.3 |
| | | L | 0 | 0 | 0 | 1.7 |
| | Gas | S | 0 | 0.2 | 0 | 0 |
| Peak Load | | | 9.4 | 11.2 | 10.1 | 33.0 |
| Export | | | 2.9 | 3.5 | 3.1 | 10.2 |

[a] S stands for small unit capacity from 0-300 GW, M stands for medium unit capacity from 300-600 GW, and L stands for large unit capacity from 600-1000 GW.

The potential renewable energy resources of the Northeast energy system are assessed using the Section 2.2 methodology, and the annual average capacity factors for wind and solar are displayed in Fig. 5, which illustrates the differences in the geographic distribution of renewable energy resources in the Northeast region. Wind energy resources are more abundant in the Northeast region and are mainly concentrated in the southwestern EIM and northwestern LN; solar energy resources are lower than wind energy resources and are centralized in the southern EIM and LN, with the overall trend decreasing with increasing latitude.

Table 4. Northeast grid inter-regional transmission line data [49]

| From region | EIM | HLJ | LN | EIM | EIM |
|---|---|---|---|---|---|
| To region | JL | JL | JL | HLJ | LN |
| Existing capacity (GW) | 3.6 | 4.8 | 4.8 | 2.4 | 7.2 |
| Expanded capacity limit (GW) | 9.6 | 9.6 | 9.6 | 9.6 | 9.6 |
| Capital cost (\$/kW) | 167.30 | 60.18 | 65.91 | 184.74 | 145.79 |
| Length (km) | 587.26 | 115.57 | 140.78 | 663.99 | 492.50 |
| Lifetime (years) | 40 | 40 | 40 | 40 | 40 |

The hourly capacity factors for wind and solar for each region are used as model input data for the renewable energy output calculations. The maximum installed potential of wind and solar energy in each region is obtained from [50] and [51].

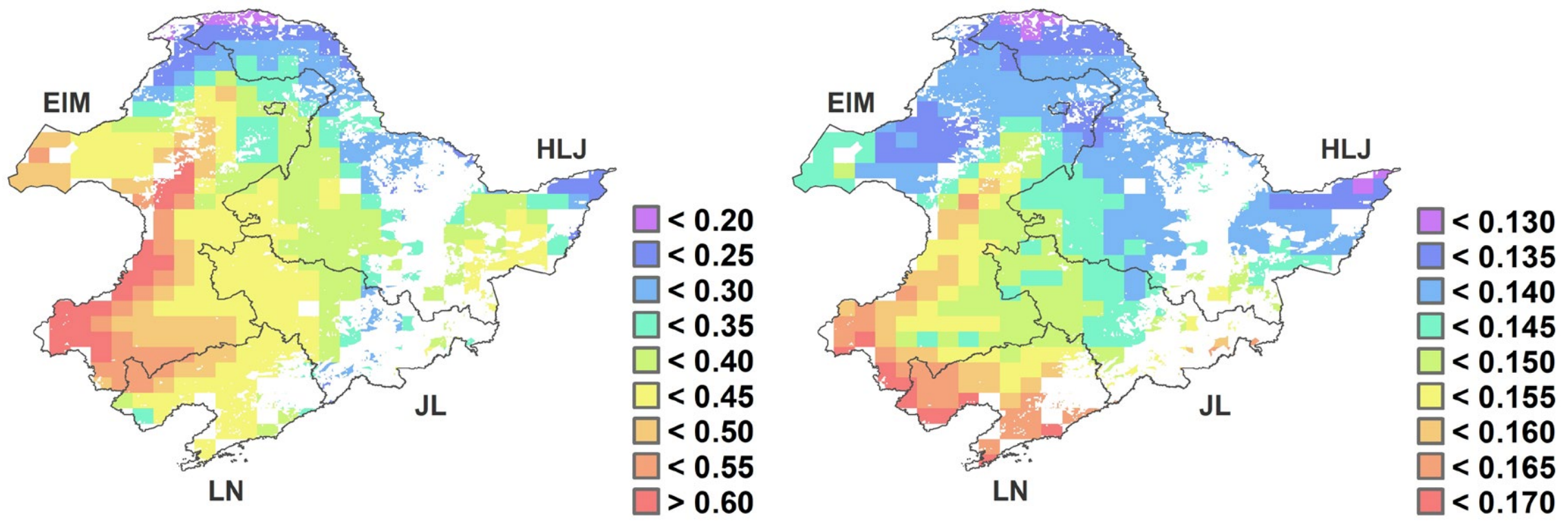

(a) Wind capacity factor (b) Solar capacity factor

Fig. 5. Renewable energy annual average capacity factors in Northeast

The model proposed in this paper is essentially a large-scale linear programming problem, which is modelled in MATLAB in conjunction with YALMIP for simulation, and Gurobi is invoked to solve the problem.

### *3.2. Scenarios setting*

Based on the above data and modelling, to explore the planning configurations and details of the roles of key equipment for multi-energy systems with a complete hydrogen energy chain in 2050, eight types of scenarios are set up, as detailed in Table 5. Each of these scenarios includes the base electricity sector's equipment (TU, WT, and PV), the heat sector's equipment (EB and HST), electric energy transmission lines, and hydrogen pipelines. The scenarios considering RPS include different levels of 20%, 40%, 60%, 80% and 100%. In addition, E, H, and $H_2$ in Table 5 denote electricity demand, thermal demand, and hydrogen demand, respectively.

Table 5. Different scenarios setting

| Scenario | Objective | RPS | Demand | | | ES | | Hydrogen energy chain | | | | |
|---|---|---|---|---|---|---|---|---|---|---|---|---|
| | | | E | H | $H_2$ | BES | HPS | EC | COP | HS | FC | HT |
| 1 | Pareto | × | √ | √ | √ | √ | √ | √ | √ | √ | √ | √ |
| 2 | Min-cost | × | √ | √ | √ | √ | √ | √ | √ | √ | √ | √ |
| 3 | Min-$CO_2$ | × | √ | √ | √ | √ | √ | √ | √ | √ | √ | √ |
| 4 | Min-cost | 20-100% | √ | √ | √ | √ | √ | √ | √ | √ | √ | √ |
| 5 | Min-cost | 100% | √ | √ | √ | √ | √ | √ | √ | √ | × | × |
| 6 | Min-cost | 100% | √ | √ | √ | √ | √ | √ | × | × | √ | √ |
| 7 | Min-cost | 100% | √ | √ | × | √ | √ | × | × | × | × | × |
| 8 | Min-cost | 100% | √ | √ | × | √ | × | × | × | × | × | × |

### *3.3. Validation of the hydrogen chain-based fast clustering optimization method*

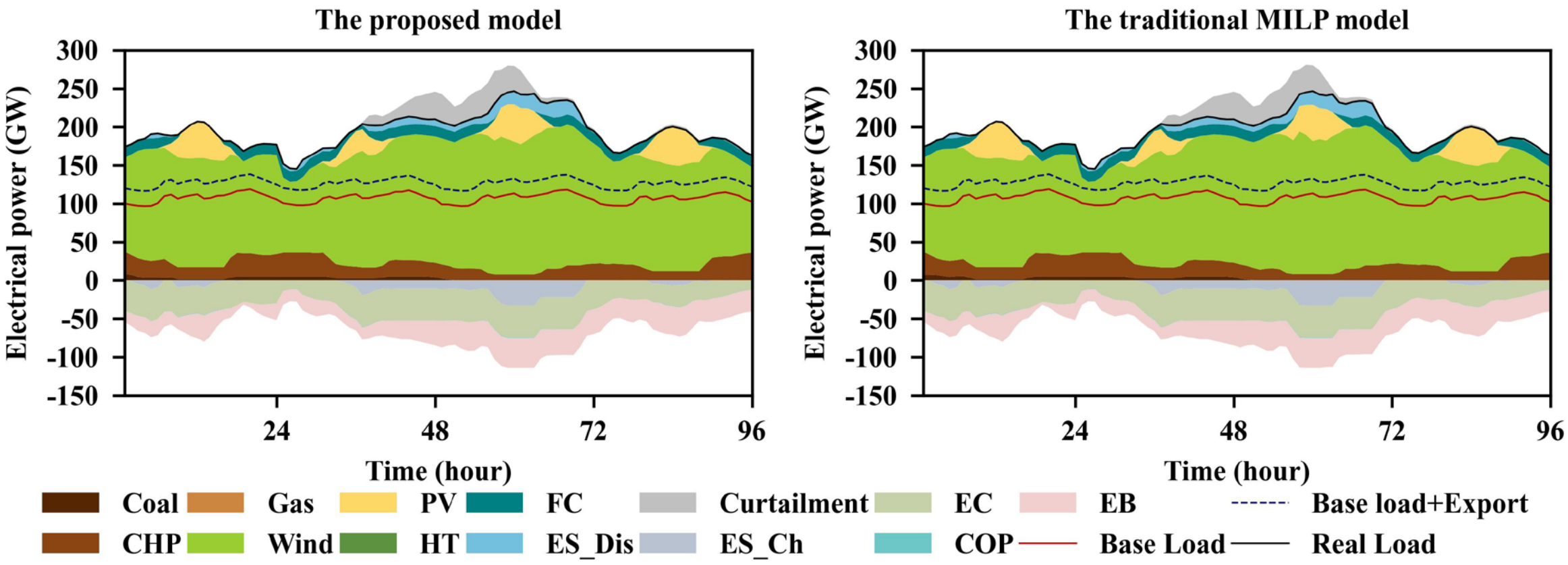

Fig. 6. Electrical power hourly balancing comparison between the proposed model and the traditional MILP model

To validate the effectiveness of the proposed hydrogen chain-based fast clustering optimization method, we simulate the consecutive 4-day scheduling of the Northeast energy system in March using the proposed model and the traditional MILP model. The traditional MILP model utilizes the traditional unit combination approach [52] to establish flexibility constraints for hydrogen energy equipment. The electricity sector hourly balanced scheduling results are shown in Fig.6. Compared with the traditional MILP model, the hourly power of the proposed model is almost the same, and the errors in power generation and power consumption of the different equipment types are less than 2%. The computation time is only 3.26 s, which is approximately 920 times faster than that of the traditional MILP model. Furthermore, as the simulation time grows, the speed advantage of solving the proposed model becomes more prominent. When the model scale and simulation time exceed a certain threshold, the traditional MILP model cannot directly solve optimization problems containing a large number of integer variables, while the model proposed in this paper can achieve efficient solution as well as save resource on the computation.

### *3.4. Pareto analysis of costs and $CO_2$ emissions*

The Pareto frontier of cost and $CO_2$ emissions for the energy system in scenario 1 is shown in Fig. 7, which illustrates the cost breakdown for different equipment. At the optimal cost, the total cost of the energy system is approximately \$23.8 billion, mainly from thermal units, and the highest system $CO_2$ emissions are approximately 1136.8 million tons. Costs rise slowly at first as $CO_2$ emissions are reduced, and it is noteworthy that a cost increases by only 30% when $CO_2$ is reduced by 60%, with the increase in costs coming mainly from newly installed WT replacing TU. For further deep decarbonization of the energy system, the cost proportions of PV, EC, and ES have started to increase, and the proportion of TU continues to decrease. As the installed WT capacity gradually reaches the regional maximum and the additional PV increases significantly, the total system cost rises more rapidly. At the optimal $CO_2$ emission, the total system cost is approximately \$72.3 billion. Meanwhile, WT and PV account for the largest proportion of 71.23%, and the energy system considers the utilization of HT and FC for clean energy supply, both of which account for a lower proportion of the cost. Other equipment costs are changed less in the system. In the early stages of carbon reduction, the revenue of the system, which comes primarily from the sale of hydrogen to meet hydrogen demand, remains essentially unchanged. When the system is deeply decarbonized, additional hydrogen is produced for hydrogen power generation, with revenues increasing as the byproduct, oxygen, increases.

The $CO_2$ emissions and $CO_2$ reduction costs of the energy system in the Pareto analysis are indicated in Fig. 8. The optimal $CO_2$ emissions scenario is selected as the baseline scenario to compare and analyze the $CO_2$ emission reductions of other typical Pareto points at the Pareto frontier. The $CO_2$ reduction cost is defined as the ratio of the incremental cost and $CO_2$ emission reduction of a typical Pareto point to the baseline scenario. When $CO_2$ emissions are reduced by 70% relative to the baseline scenario, the $CO_2$ reduction cost is $14.1/ton, and when the system achieves zero-carbon emissions, the $CO_2$ reduction cost is $42.7/ton. According to [53], the social cost of carbon (SCC) in 2050 under a 3% discount rate is estimated to be $44.6/ton, which is higher than the $CO_2$ reduction cost for all typical Pareto points. This indicates that the cost induced by achieving zero-carbon emissions in the future using the proposed multi-energy coupled energy system with the complete hydrogen energy chain is within the socially acceptable range.

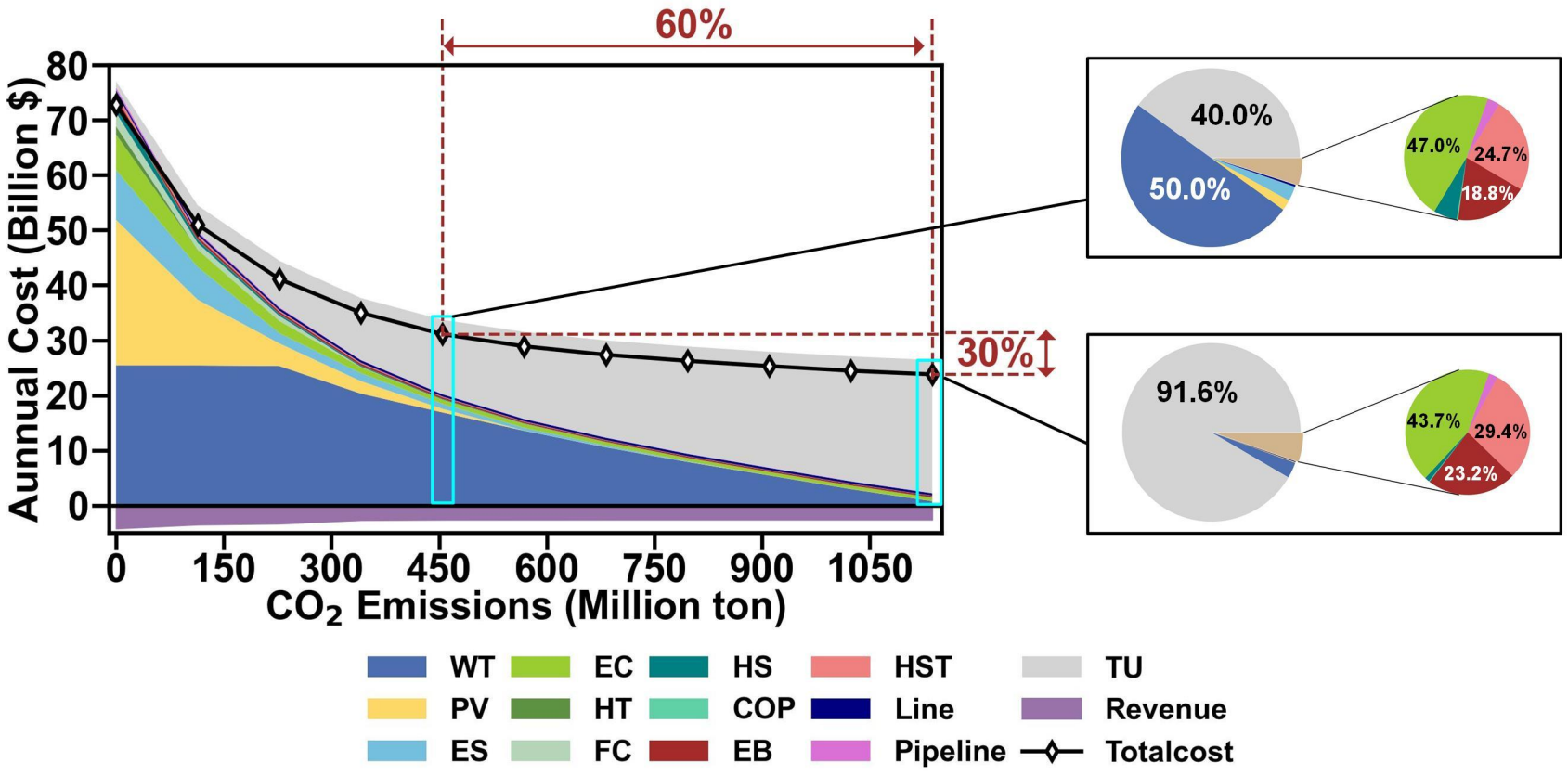

Fig. 7. Cost-emission Pareto frontier in scenario 1 (the corresponding values are marked where the percentage value more than 10% in the pie chart)

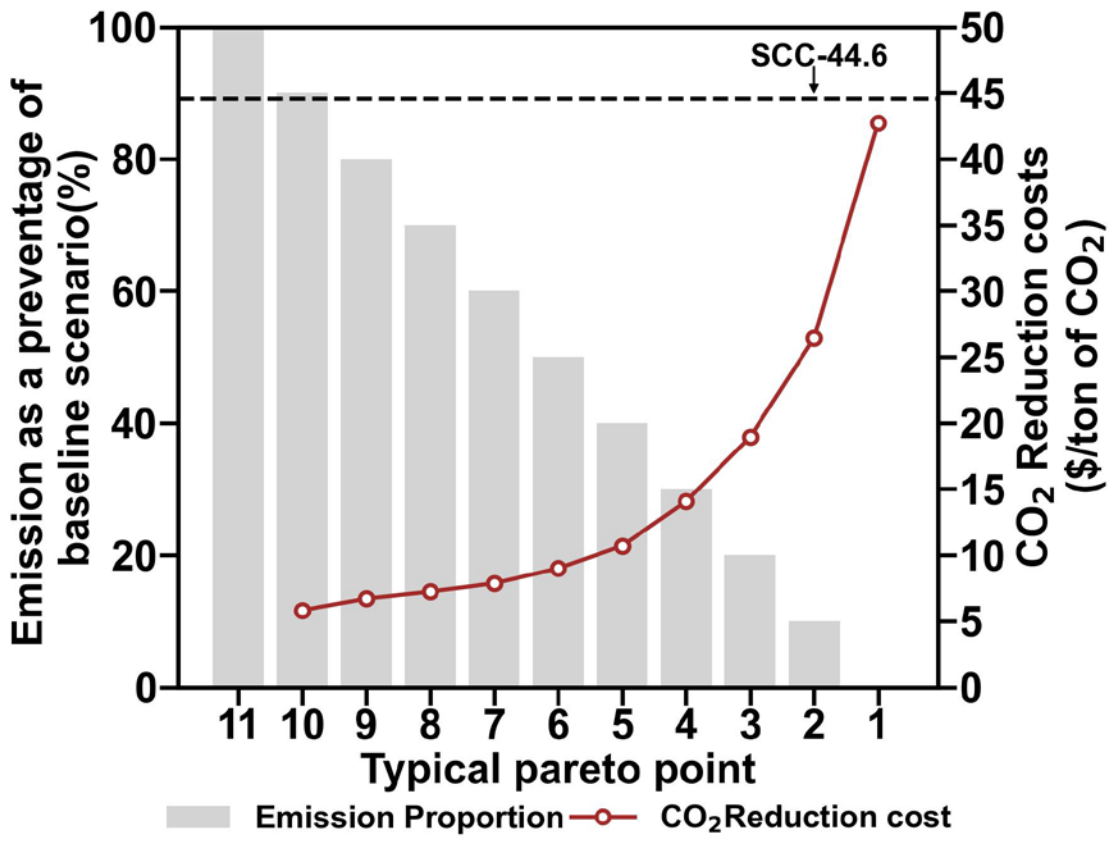

Fig. 8. $CO_2$ reduction costs in Pareto analysis in scenario 1

### *3.5. Transmission network analysis*

To visualize the inter-regional transmission network expansion, in this section, electric transmission lines and hydrogen pipelines are shown in the electricity and hydrogen sectors, respectively, and heat energy is not considered for inter-regional exchange [29]. The installed base equipment in the electricity, heat, and hydrogen sectors is represented in blue, red, and green in Fig. 9, which is summarized in scenario 2 (min-cost) and scenario 3 (min-$CO_2$).

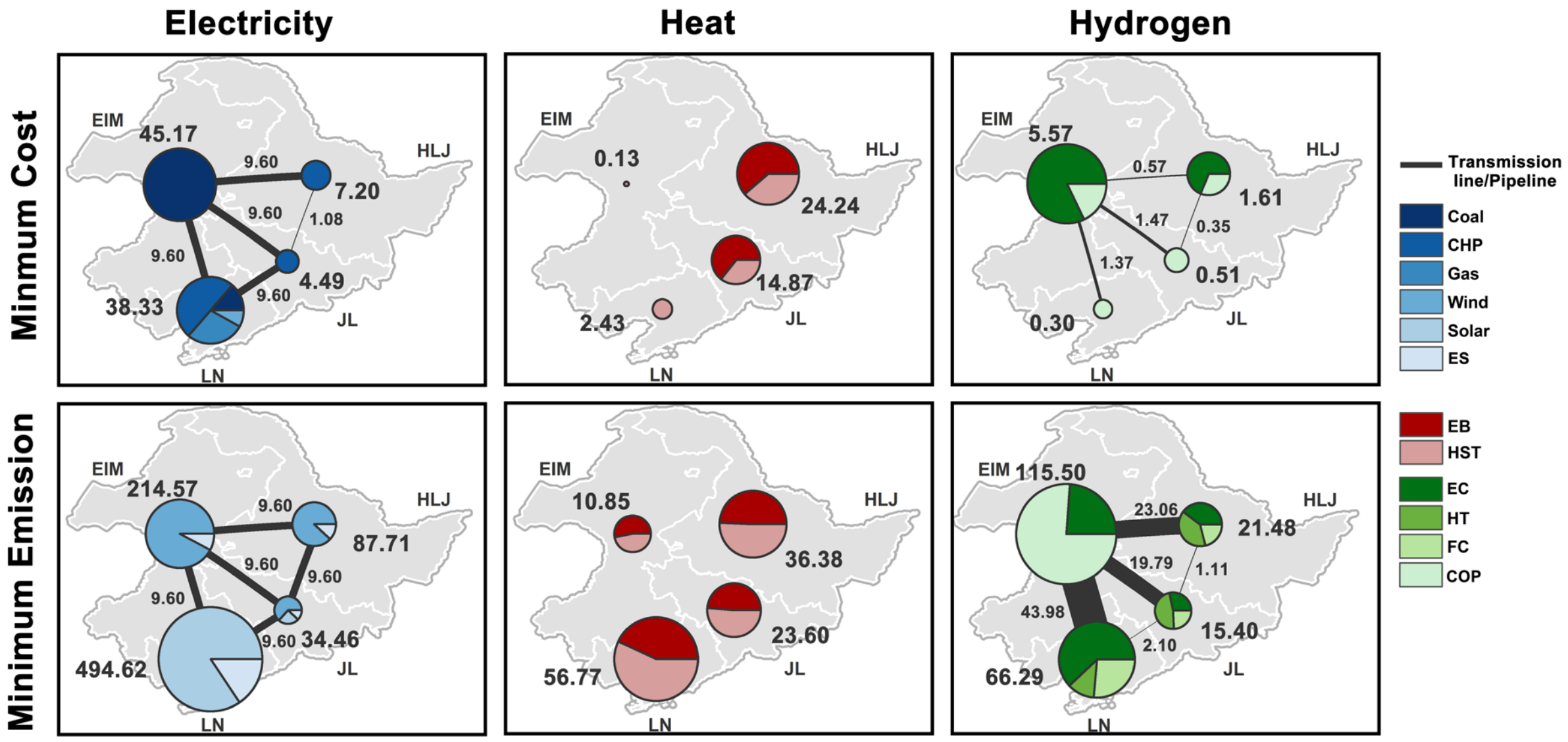

Fig. 9. Geographic distribution of transmission lines and installed capacity in the min-cost and min-$CO_2$ scenarios

Table 6. Hydrogen transmission cost

| | Hydrogen pipeline | |
|---|---|---|
| Transmission cost ($/km·GWh) | 2.484[54] | |
| | Min-$CO_2$ | Min-cost |
| Transmission cost (Million $/year) | 378.0 | 32.8 |
| Share of total cost (%) | 0.52 | 0.14 |

Comparing the min-cost and min-$CO_2$ scenarios, the biggest difference in the electricity transmission network is that the capacity of the HLJ-JL transmission line changes from 1.08 to 9.60 GW, with all other transmission lines reaching their capacity limits. This indicates that the expansion of power transmission lines in the Northeast energy system is necessary to help optimize regional resource allocation and improve the flexibility and economy of the energy system. In the min-cost scenario, the EC prioritizes investments in regions with more installed capacity of renewable energy units and transports extra hydrogen energy between regions through pipelines. This hydrogen is used only to meet the hydrogen load demand, and in combination with HS and COP, only a few new low-capacity hydrogen pipelines are needed to achieve a regional balance between hydrogen supply and demand. In the min-$CO_2$ scenario, HT and FC consume hydrogen for clean energy supply, and the expansion of the hydrogen pipelines in the energy system increases significantly compared with the min-cost scenario, up to an EIM-LN pipeline capacity of 43.98 GW. This suggests that the expansion of hydrogen transmission pipelines is essential to economically achieve zero carbon emissions in the future. In the process of balancing regional hydrogen energy supply and demand, EIM is mainly responsible for hydrogen production, storage, and coordination of hydrogen energy transmission. As shown in Table 6, the cost of hydrogen transmission for both the min-$CO_2$ and min-cost scenarios is less than 1%, which indicates that the coordinated planning and post-processing optimized hydrogen transmission expansion method used in this paper is reasonable.

### *3.6. Analysis of optimal generation mix under different RPS scenarios*

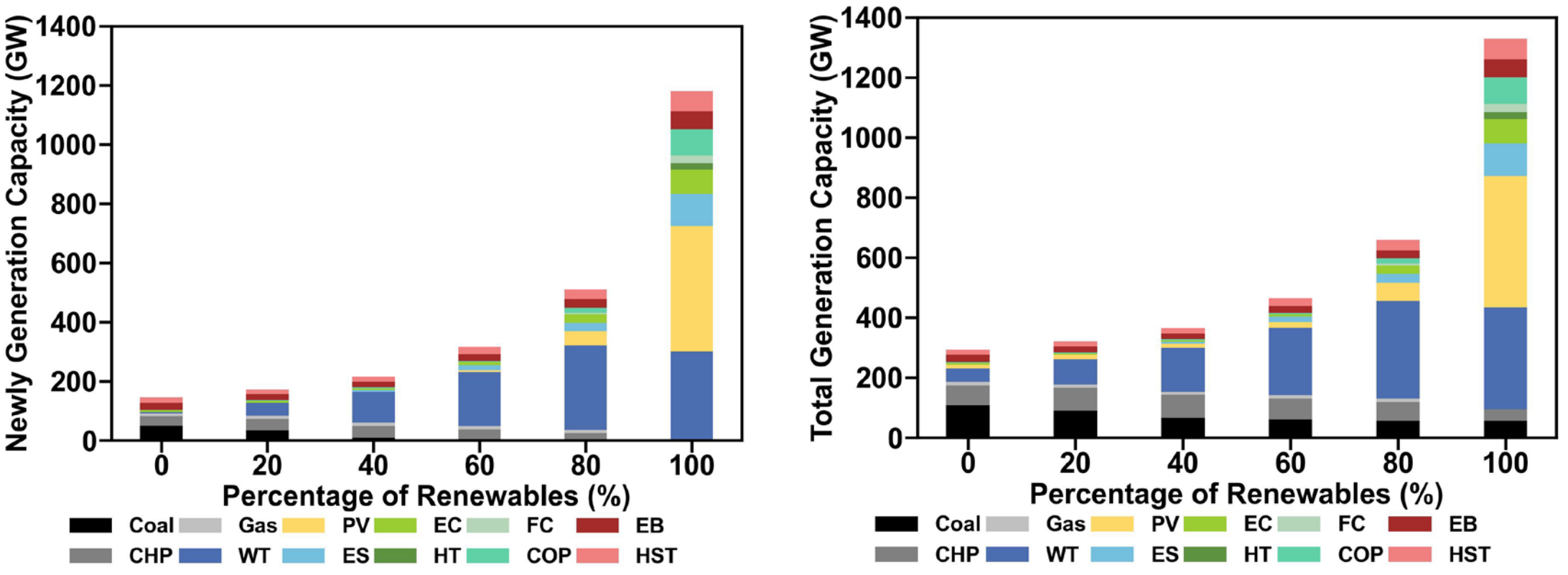

(a) Newly installed generation mix

(b) Total generation mix

Fig. 10. Newly installed and total generation mix of energy system under different RPSs in scenario 4

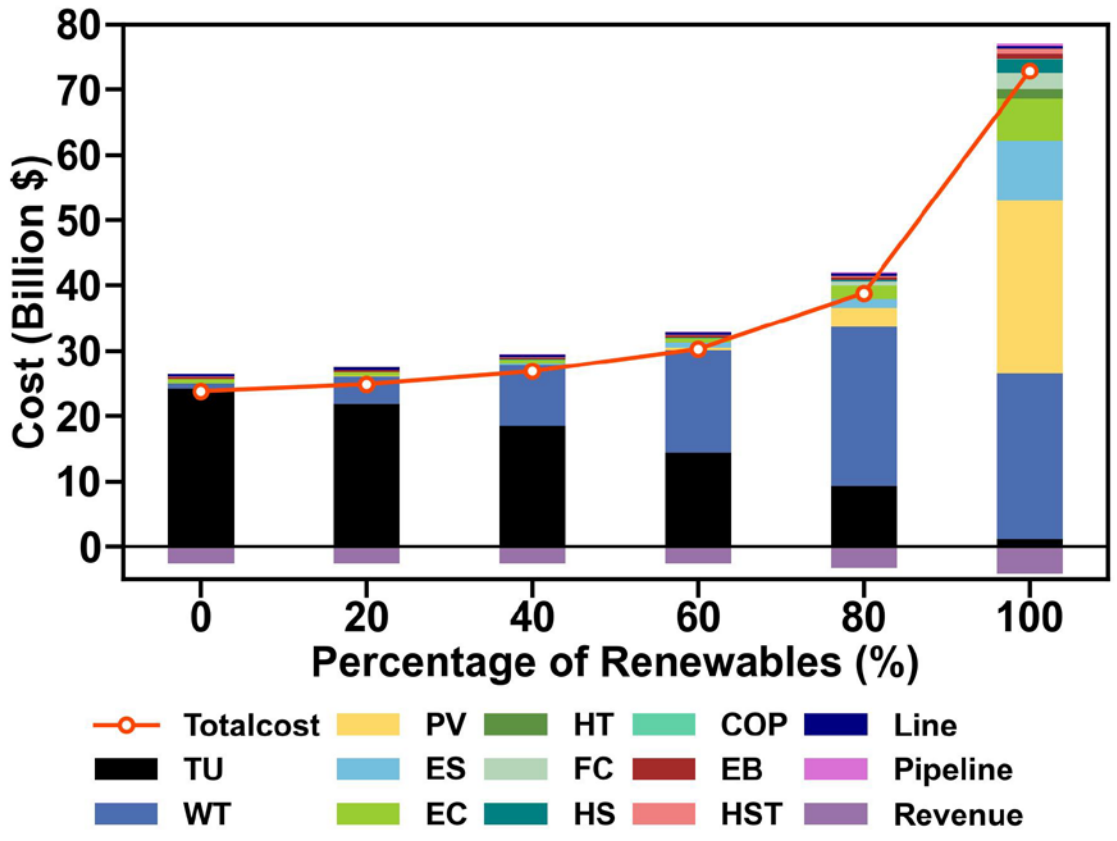

Fig. 11. Cost mix of energy system under different RPSs in scenario 4

For scenario 4, Fig. 10 illustrates the mix of newly installed and total installed capacity in the Northeast energy system in 2050 under different RPSs, while Fig. 10 presents the cost mix. With the increase in RPS, the capacity of TU gradually decreases, and the capacity of WT, PV, and ES gradually increases. The new installed capacity of CHP units increases and then decreases, mainly replacing each other with EB, and the HST plays the role of regulating as a flexible resource of heat energy and increases significantly from 80% RPS. At 100% RPS, the EB completely replaces the CHP units to become the main energy supply equipment in the heat energy sector. Combined with the waste heat utilization of hydrogen energy equipment, the EB completely realizes the clean heat supply in the Northeast energy system.

When the RPS is less than or equal to 60%, the energy system considers hydrogen only to meet hydrogen demand, with little change in EC and COP capacity. At 80% RPS, the energy system prefers to invest in more efficient FC first, even though the cost of HT is lower. To achieve the target of 100% RPS, the energy system adds approximately 423.80 GW of PV and 299.77 GW of WT, which yield significant cost additions. At this point, the energy system chooses to build both the price-competitive HT and the efficiency-competitive FC for hydrogen power generation.

As illustrated in Figure 11, as the RPS increases, the costs gradually shift from TU costs being the largest share to renewable energy costs dominating. Hydrogen energy chain-related equipment causes a relatively small increase in cost, accounting for approximately 18% of the total cost when the RPS is 100% and generating approximately 6% revenue.

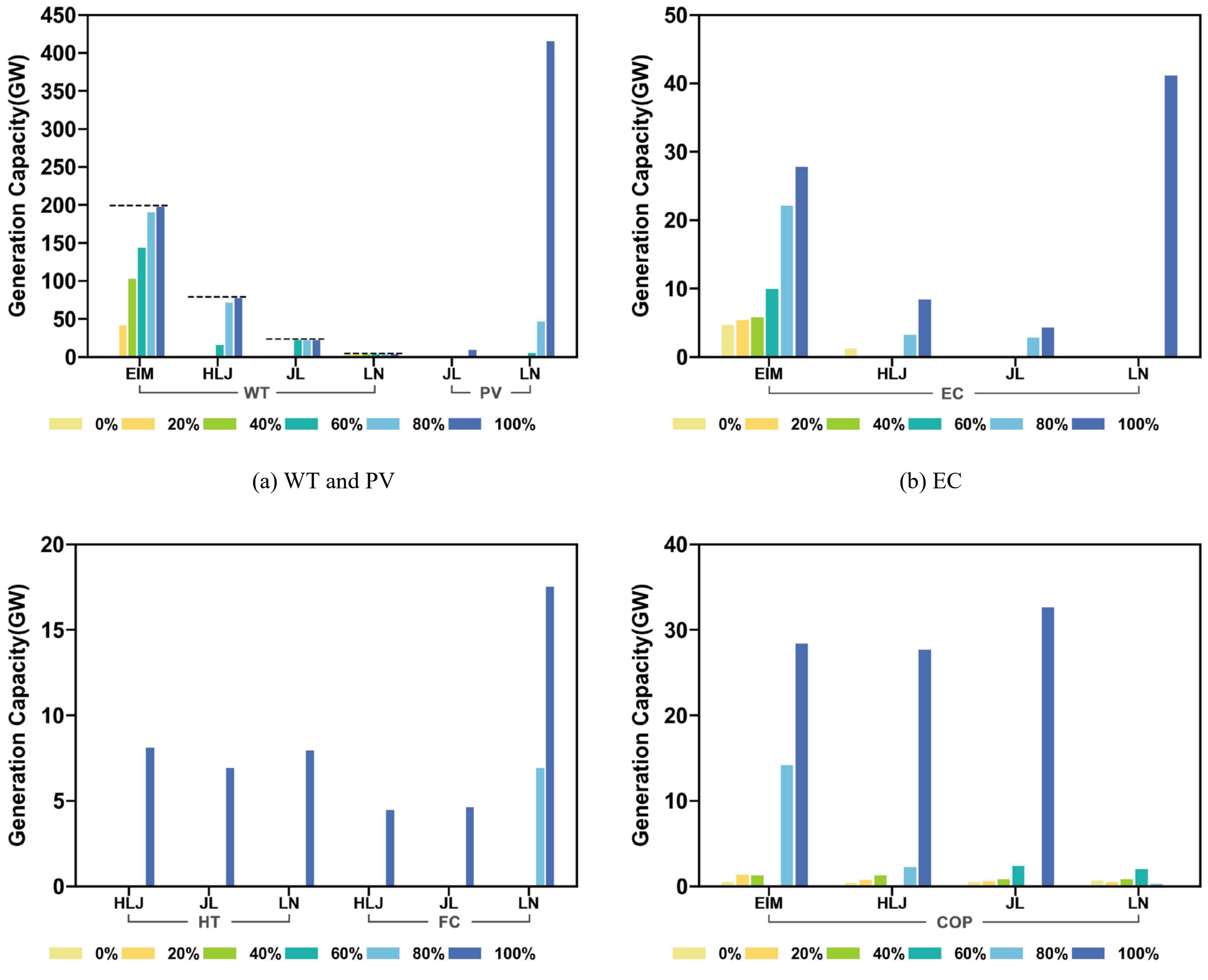

(a) WT and PV

(b) EC

(c) HT and FC

(d) COP

Fig. 12. Geographical distribution of newly installed generation under different RPSs in scenario 4

Fig. 12 illustrates the geographic distribution of the installed capacity of WT, PV, and hydrogen-related equipment at different RPSs. The Northeast energy system is more abundant in wind than solar, prioritizing investment in WT in wind-rich areas such as LN and EIM as the RPS target increases. WT investments are more evenly distributed geographically starting at 60% RPS, given the capacity limitations and the challenge of flexibility posed by high penetration renewable energy. Given that the total installed WT capacity is limited to 14.07 GW and solar resources are relatively abundant, LN started investing in PV units from 60% RPS and installed approximately 415.00 GW of capacity at 100% RPS.

The change in installed capacity and geographic distribution of EC is closely related to the distribution of renewable energy unit capacity. Before RPS reaches 60%, the total capacity of EC does not change much and is concentrated in regions with a high total installed renewable energy capacity, as hydrogen generation is not considered. The installed capacity of the COP is less variable and more evenly distributed geographically. When RPS is higher than 80%, FC with high efficiency is newly built in the LN area for energy supply due to low installed WT capacity and diurnal variation of solar power. When the RPS reaches 100%, investment in HT and FC will be needed in all regions except the EIM, where turbines are installed in abundance. The installed capacity of EC and COP begins to increase significantly to meet hydrogen generation. The integrated utilization of the complete hydrogen energy chain in the energy system can facilitate the consumption of renewable energy and improve system dispatch flexibility, contributing to the goal of a clean energy supply.

### *3.7. Selection and analysis of equipment related to hydrogen energy chain*

#### *3.7.1. Comparison of equipment performance indicators*

Table 1 summarizes key data on different technology types of equipment in a typical complete hydrogen energy chain. To better visualize the differences between various technologies and to generalize the portfolio selection laws for future energy system planning, this paper divides the above data into three categories of indicators: economy, flexibility, and efficiency. The economy is represented by capital cost, flexibility by ramping capacity, and efficiency by electrical efficiency. Since there are only cost differences in HS, the technical indicators of EC, HT, and FC are demonstrated comparatively in the form of radar charts, as shown in Fig. 13.

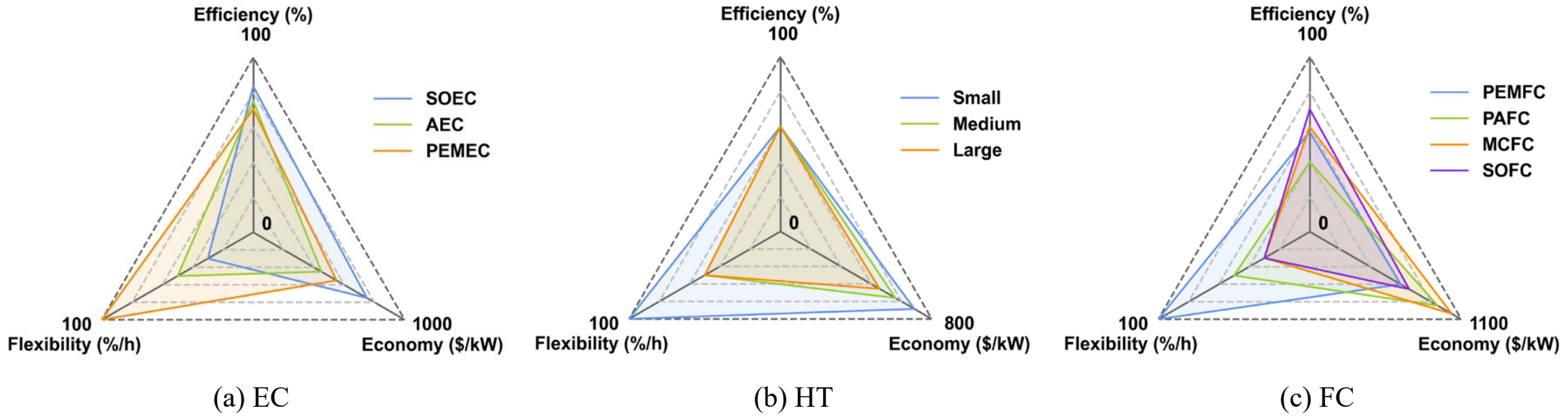

Fig. 13. Comparison between the technical indicators of different types of EC, HT, and FC

From the comparison in Fig. 12, it is concluded that PEMEC is optimal in flexibility but lowest in efficiency, AEC is optimal in economy, medium in efficiency and flexibility, and SOEC is optimal in efficiency but worst in cost and flexibility. For HT, the efficiencies of the different-sized units are consistent, with the economies of the sizes increasing in descending order from small to large. The flexibility of the medium and large units is consistent and lower than that of the small units. In terms of FC, PEMFC is optimal in terms of flexibility and economy, SOFC is optimal in terms of efficiency, while PAFC and MCFC do not stand out in terms of performance comparatively and are less economical. Belonging to the same application link of hydrogen power generation, the different technologies of HT and FC are compared with each other, with large HT having the optimal cost, SOFC having the optimal efficiency, and small HT and PEMFC having the optimal flexibility.

*3.7.2. Portfolio selection results*

From the simulation results in scenario 4, the selection of HS and HT under different RPSs tends to favor the model with a greater cost advantage, such as salt caverns for HS and L size for HT, while EC and FC have different portfolio selection results under different RPSs. We choose representative EIM and LN regions for the selection analysis of related equipment, and the results are shown in Fig. 14.

For EC portfolio selection, at RPSs below 40%, the installed capacity requirement of EC is low, inducing a cost of less than $520 million, the impact of the cost difference between models is small, and the system tends to invest in SOEC to utilize renewable energy efficiently. At 60% and 80% RPS, the installed capacity requirement of EC increases, the cost impact on the system is large, and the installed capacity of renewable energy is increased. Therefore, the AEC is sufficient to satisfy the hydrogen production demand with a cost savings of 52.65% compared to the equivalent capacity of the SOEC. When RPS reaches 100%, the system again favors the SOEC with optimal efficiency to meet the increased demand for hydrogen production due to hydrogen power generation, as the installed capacity of WT in the EIM region reaches the upper limit.

For FC portfolio selection, when the RPS is 80%, the overall installed renewable energy capacity in the LN region is only 65.15 GW, and the system favors SOFC with optimal efficiency to supplement the energy supply gap. To achieve 100% RPS, a large number of new PV are installed in LN, and there is a large difference in PV output between daytime and nighttime, so the system selects the PEMFC with optimal flexibility.

From the above results, it can be summarized that the portfolio selection law for EC is as follows: when the system has little renewable energy surplus and little EC installed demand, the SOEC with optimal efficiency is preferred; when the system has much renewable energy surplus and large EC installed demand, the AEC with the optimal cost is preferred. The portfolio selection law for FC is that according to the demand for energy supply, the SOFC with optimal efficiency or the PEMFC with optimal flexibility and cost is preferred. Furthermore, the combination of different models according to the actual situation of the energy system can realize the efficient, flexible, and economic use of energy.

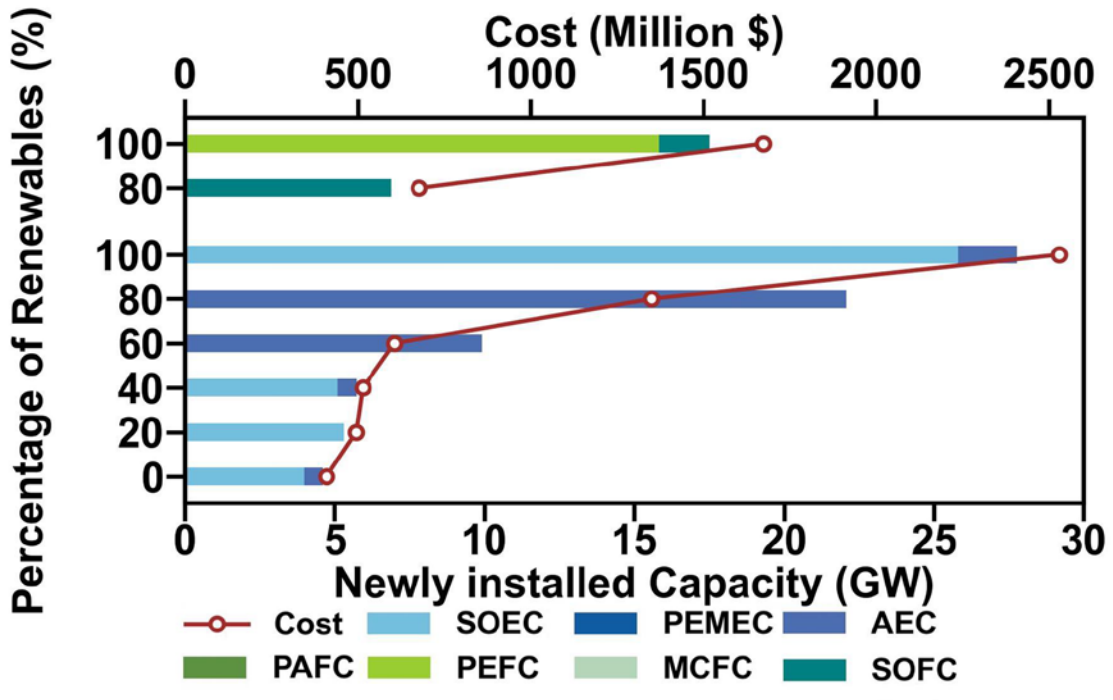

Fig. 14. Portfolio selection of EC and FC with different RPSs in scenario 4

*3.7.3. Analysis of hydrogen energy chain configuration results*

Table 7 illustrates the total cost and renewable energy curtailment results for the different scenarios, where scenario 4 contains the complete hydrogen energy chain, scenario 5 ignores the hydrogen power generation link, scenario 6 ignores the hydrogen storage link, scenario 7 ignores the complete hydrogen energy chain and hydrogen demand, and scenario 8 without the HPS is based on scenario 7.

Table 7. Costs and renewable energy curtailment under different configurations

| | Scenario 4 | Scenario 5 | Scenario 6 | Scenario 7 | Scenario 8 |
|---|---|---|---|---|---|
| Total cost (Billion $) | 72.77 | 910.79 | 1026.86 | 911.12 | / |
| Renewable energy curtailment (GWh) | 6821.45 | 109996.45 | 21376.45 | 228787.54 | / |
| Renewable energy curtailment rate (%) | 0.39% | 5.87% | 1.15% | 12.21% | / |

In contrast to scenario 4, the cost of scenarios 5, 6, and 7 increases a great deal, and scenario 6 in particular costs 14 times more than scenario 4. All three cost increases come primarily from HPS replacing salt cavern hydrogen storage to meet the seasonal supply/demand balance of the energy system. Disregarding hydrogen and HPS with long-term storage characteristics, scenario 8 is unsolved at 100% RPS. Comparing the renewable energy curtailment, the order from high to low is scenario 7, scenario 6, scenario 5, and scenario 4, which indicates that more renewable energy can be consumed at 100% RPS despite the lower energy conversion efficiency of the hydrogen energy generation link. Compared to scenario 7, scenario 4 has 97.0% less renewable energy curtailment and 11.82% lower curtailment rate. The above results demonstrate that inexpensive salt cavern storage and hydrogen power generation technologies contribute to the seasonal supply-demand balance and renewable energy consumption of the energy system. The integrity of the hydrogen energy chain is essential for balancing the economics and energy utilization of the energy system under 100% RPS.

### *3.8. Seasonal analysis of operation and hydrogen storage*

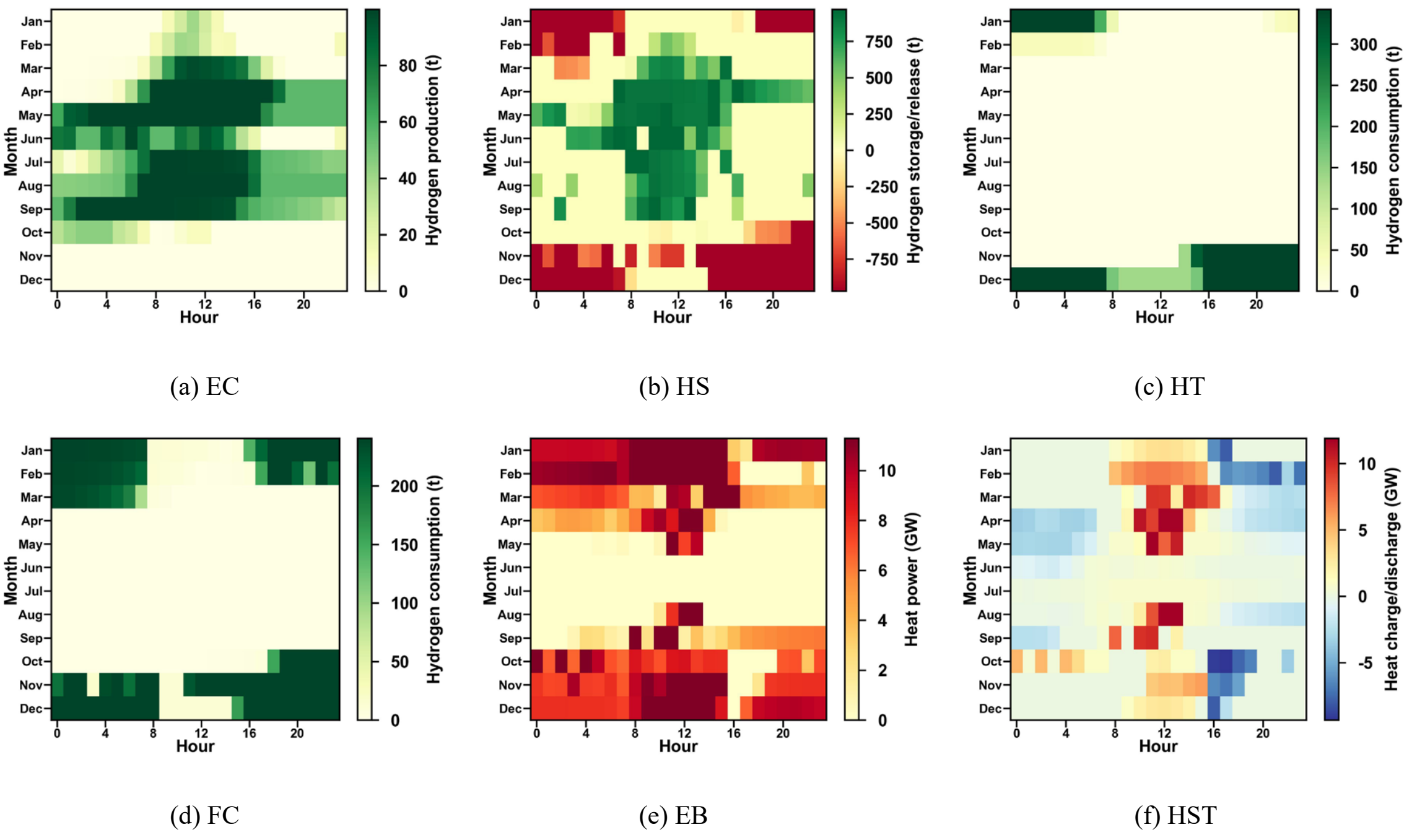

Fig. 15. Hourly operation of different equipment in JL on a typical day of the month throughout the year in scenario 4

Fig. 15 illustrates the hourly operation of the hydrogen energy chain-related equipment, EB, and HST in JL on a typical day of the month throughout the year. The horizontal axis represents the hourly sequence in a day, the vertical axis represents the monthly sequence in a year, and the color bars indicate the corresponding values through different colors and shades. The hydrogen sector and the thermal sector have large operational differences between the different seasons.

There is a very strong correlation between the operation of the hydrogen energy chain-related equipment. As shown in Fig. 15-(a), the hydrogen production of EC is concentrated in spring and summer, and during the day it is concentrated in the daytime when the PV output is high and part of the day when the turbine output is high, which is consistent with the HS storage period in Fig. 15-(b). The hydrogen release period of HS is concentrated in the nighttime of winter, which is consistent with the consumption of hydrogen by the HT and the FC operating synergistically in Fig. 15-(c) and (d). The waste heat utilization of EC, HT, and FC connects the hydrogen and thermal sectors, and the thermal power of EB is concentrated in heating periods such as early spring, late fall, and winter as shown in Fig. 15-(e). There are seasonal complementarities in the operating periods of EB and EC. In addition, as shown in Fig. 15-(f), the heat generated by the EB at midday when the PV output is high is stored in the HST and released in the evening and morning. The energy system realizes the coupling of electricity, heat, and hydrogen through a complete hydrogen energy chain, and the coordinated operation of each piece of equipment realizes the efficient consumption and optimal allocation of renewable energy between seasons and within days.

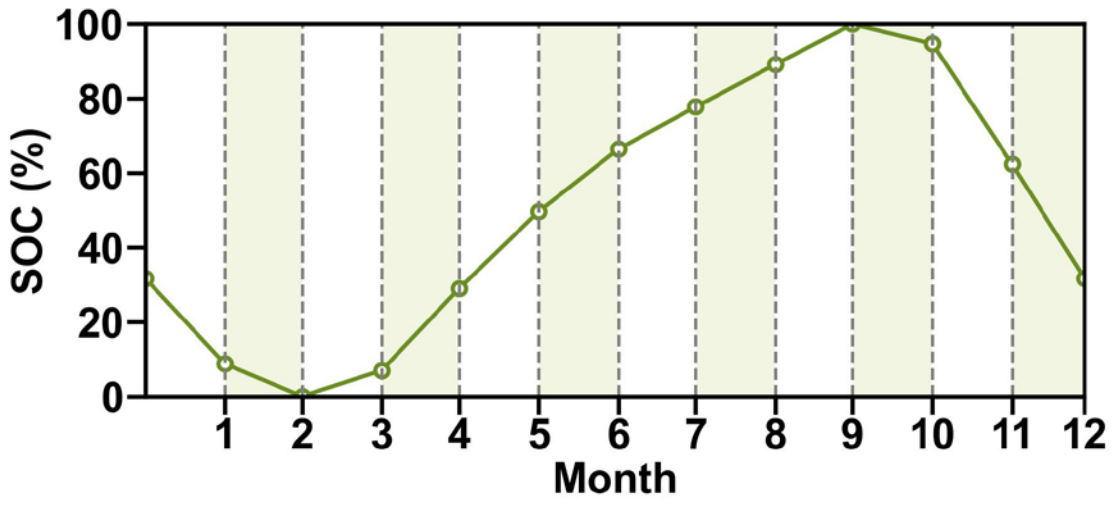

Fig. 16. Variation of SOC for hydrogen storage in salt caverns in Jilin over the course of a year in scenario 4

Fig. 16 shows the SOC variation in salt cavern hydrogen storage in JL throughout the year, with cumulative changes for each month, and the stored hydrogen reaches its peak in September.

### *3.9. Sensitivity analysis*

Taking into account uncertainties such as technological advances and price fluctuations, we conduct a sensitivity analysis of key cost parameters using scenario 4 as a baseline: a) EC cost change by ±30%; b) HT cost change by ±30%; and c) FC cost change by ±30%. The variation in the total system cost and hydrogen energy chain equipment installed capacity compared to the baseline scenario is summarized in Fig. 17. As demonstrated, total cost, HS, EC, ES, and PV are insensitive to changes in cost parameters, with variations of less than 8%, contrary to HT and FC, where the cost parameter has a significant impact on the installed capacity, even to decide whether the system will be installed with only HT or FC. Since the overall efficiency of SOFC and PEMFC is higher than that of HT, only less installed capacity of FC is needed to replace HT when the cost changes. Considering the factors of efficiency, cost, and NOx emission, we regard FC as having broader development prospects in the future hydrogen power generation field.

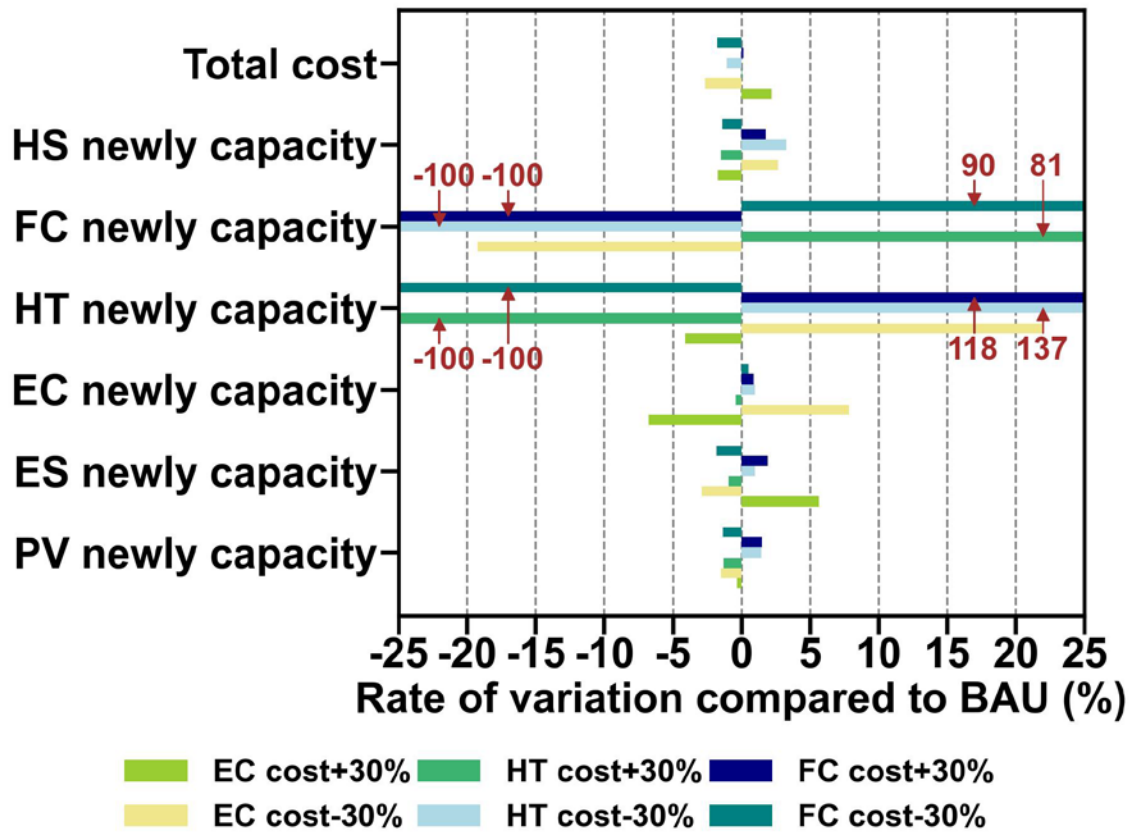

Fig. 17. Sensitivity analysis under changes in key cost parameters

## 4. Conclusion

As hydrogen energy is gaining more attention, a comprehensive quantitative analysis of the entire hydrogen energy chain is significant for energy system carbon-neutral planning. This paper proposes a high-resolution electricity-thermal-hydrogen-coupled energy system collaborative planning model incorporating the spatiotemporal distribution of renewable energy sources, which enables spatial geographic resource allocation and temporal operation optimization. The model incorporates a fast clustering optimization method to establish a complete hydrogen energy chain including production, compression, storage, transportation, and application, taking into account operational flexibility, hydrogen pipeline planning, and equipment waste heat utilization. This paper additionally analyzes the portfolio selection of equipment types related to the complete hydrogen energy chain based on performance indicators such as economy, system flexibility, and energy efficiency, and examines the details of the roles of the related equipment in the system as well as the coupling relationship between energy flows.

From the Pareto optimization results, the energy system can reduce carbon emissions by 60% with a 30% increase in cost and achieve zero carbon emissions when the carbon reduction cost is $42.7/ton lower than that of the SCC. This suggests that the model can provide economically feasible decarbonization solutions for the energy system in the future. In addition, the expansion of electric transmission lines and hydrogen pipelines can facilitate inter-regional energy flow under high renewable energy penetration, which is essential for the optimization of regional resource geographic allocation.

By setting up scenarios with different RPSs, this paper also simulates the variation in the geographic distribution of capacity investment. The geographic configuration of EC capacity tends to favor regions with more installed capacity of renewable energy units, facilitating the utilization of surplus renewable energy sources for hydrogen production. The capacity of EC varies slightly at less than 60% RPS, and FC and HT start to be added at 80% and 100% RPS, respectively. With the increase in hydrogen consumption, the installed capacity of EC appears to increase significantly. At different RPSs, both HS and HT tend to be the types with the optimal investment cost, while EC and FC demonstrate different portfolio selection results. The combination of different equipment types such as AEC, SOEC, PEMFC, and SOFC can be operated in a combination that takes into account the economic, flexibility, and efficiency needs of the energy system. The absence of selected links in the hydrogen energy chain configuration, such as the hydrogen storage link and the hydrogen power generation link, can lead to a rapid increase in costs and renewable energy curtailment.

In addition, this paper discusses in detail the full-year operation of the various equipment in the hydrogen and thermal sectors under 100% RPS, where the energy systems exhibit distinct seasonal characteristics. During the heating period, the surplus renewable energy generation is mainly consumed by the EB, while in other periods it is consumed by the EC to produce hydrogen. The surplus hydrogen is stored in the HS and used for nighttime energy supply in winter by hydrogen generation equipment such as HT and FC. The complete hydrogen energy chain can promote the coupling and synergistic operation of electricity, heat, and hydrogen to achieve the efficient consumption of renewable energy, in which the utilization of pipelines for inter-regional transportation and inexpensive salt caverns for seasonal storage is crucial to achieving the optimization of resource allocation in the energy system on a long-term scale.

According to the sensitivity analysis results, the future fluctuation of equipment cost has a large impact on the capacity of HT and FC, and the development of cost reduction and technical performance improvement is essential for the future competition between the two. The flexibility of hydrogen demand is not considered in this paper, so hydrogen consumption by hydrogen fuel cell vehicles will be considered in future work. In addition to hydrogen pipelines, exploring the role of hydrogen transportation modes such as pipe trailers and liquid hydrogen tankers in energy systems is significant for optimizing energy allocation and needs to be further investigated in the future.

## Acknowledgements

This work was supported by National Natural Science Foundation of China-Enterprise Innovation and Development Joint Fund (U22B20102).